\documentstyle[12pt]{article}
\begin{document}
\author{\normalsize\bf Yu.A. Markov$\!\,$\thanks{e-mail:markov@icc.ru}
$\,$and M.A. Markova$^*$}
\title{Nonlinear dynamics of soft boson excitations\\
in hot QCD plasma I: plasmon-plasmon scattering}
\date{\it Institute of System Dynamics\\
and Control Theory Siberian Branch\\
of Academy of Sciences of Russia,\\
P.O. Box 1233, 664033 Irkutsk, Russia}

\thispagestyle{empty}
\maketitle{}


\def\theequation{\arabic{section}.\arabic{equation}}

\[
{\bf Abstract}
\]
On the basis of pure gauge sector of Blaizot-Iancu equation, we derive
kinetic equation of Boltzmann type, taking into account $2n+2$-colorless
plasmon decay processes, $n=1,2,\dots\,.$ Using so-called Tsytovich
correspondence principle, a direct connection between matrix elements of the
plasmon decay processes and certain effective current, generating these processes,
is established. The procedure of calculation of matrix element for simplest
four-plasmon decay is considered comprehensively. The limiting value of the
plasmon occupation number ($\sim 1/g^2$, where $g$ is a strong coupling)
wherein all plasmon decays with $n\geq 1$,
contribute to the right-hand side of the Boltzmann equation, is defined. The
iterative method of calculation of matrix elements for higher decay
processes ($n>1$), is proposed, and a problem of their gauge-invariance is discussed.
Proceed from the general reasons the problem of extension of suggested approach
to the case of color plasmons, is considered. The explicit form of linearized
Boltzmann equation for color plasmons is written out, and it is shown that this
equation covariantly conserves the color current, resulting from color-plasmon
number density.


\newpage

\section{Introduction}
\setcounter{equation}{0}

This work consisting of three parts, is concerned with in-depth analysis of
dynamics of boson excitations in hot QCD plasma at the soft momentum scale
($\sim gT$, where $T$ is a temperature of a system), within the hard thermal
loop (HTL) effective theory. Although this range of momentum was studied
in detail in the literature (see review of Blaizot and Iancu \cite{bla1}),
but nevertheless the dynamical processes of higher order in the coupling
$g$ connected with high-loop diagrams of an effective perturbation theory
\cite{bra1}, remain to be practical analyzed.
To such processes in particular we carry over the
processes of the following type: the scattering of soft modes among themselves and
the processes of scattering of two and more plasma waves by hard thermal
particle. This and accompaying papers are concerned with research of the processes
within a real-time formalism based on kinetic equation for soft excitations.
For sufficiently high energy level of the soft plasma excitations (exact
estimations will be given below) these higher processes of interaction
play an important role identical to simplest scattering process including one
soft collective quantum and hard particle, and is called nonlinear Landau damping.
In the case of non-Abelian plasma, for boson excitations, the nonlinear
Landau damping was studied in
detail in Ref.\,\cite{mar1}. In \cite{mar1} it was shown that the nonlinear
Landau damping rate is closely related to the damping rate obtained in
Ref.\,\cite{bra2} in one-loop approximation of an effective perturbation theory.
In the second part of our paper \cite{mar2} the extension of an approach developed
in Ref.\,\cite{mar1} to the case of scattering of arbitrary number of waves by
hard thermal particle will be introduced, and exact estimations for oscillation
amplitudes, wherein these higher processes are of the same order in $g$,
will be given. The application of the theory developed there to the problem
of energy loss of a color charged particle traversing the hot QCD plasma
will be considered. In the third part the plasmon production
by bremsstrahlung and processes of multiple scattering with emission of
soft boson waves (longitudinal or transverse) will be studied,
what can be useful for research of
Landau-Pomeranchuk-Migdal effect in hot gauge plasma.

Here, in Paper I we have restricted ourselves to study of
the processes connected with self-interaction of the soft excitations. The
study of these processes is to be preceded by consideration of more
complicated scattering processes including soft modes and hard particles
which contain self-interaction of the soft modes as a required constituent
element.

In our previous paper \cite{mar3} a first attempt of research of dynamical
process of nonlinear self-interaction of longitudinal colorless excitations
(colorless plasmons) based on the Boltzmann equation for plasmon number
density $N^l$, was presented. The process of elastic scattering of two colorless
plasmons for which the appropriate scattering probability was calculated, is
considered. Unfortunately, method of derivation of collision term proposed
there, is rather cumbersome and gives no way of its direct extension to
calculation of collision term taking into account the scattering processes
involved arbitrary (even) number of colorless plasmons. The present work is
based on use of some different approach allowing immediately to reduce the
problem of calculation of the collision term to the problem of calculation
of the scattering probabilities for appropriate scattering processes.
Above-mentioned assertion that for certain energy level of the soft
excitations all processes of self-interaction of the plasma collective modes
are of the same order in $g$, is stated more concrete in terms of the plasmon
number density, and the equation of the Boltzmann type defines its
space-time evolution. In present paper it shown that with the condition when
the plasmon
occupation number is of order $1/g^2$, the right-hand side of the Boltzmann
equation represents a sum of an infinity number of terms having a sence of
$(2n+2)$-plasmon collision integrals, $n = 1, 2,\ldots\,.$ The last circumstance
is in reality a different restatement of well-known \cite{efr, bla2} very
nontrivial fact that for strong field $A_{\mu}(X)\sim T$, all $N$-gluon
amplitudes with $N\geq 2$ contribute to the induced current
$j_{\mu}^{\,{\rm ind}}[A]$ at the same order in the coupling. The HTL-amplitudes
here represent coefficient functions in integrands for
functional expansion of $j_{\mu}^{\,{\rm ind}}[A]$ in powers of gauge field
$A_{\mu}$.
In this paper by introducing notions of interacting and free
soft gauge fields, we somewhat deepen analysis of this circumstance.
In particular, it shown that the induced current has the more physical content
if it will be represented as an expansion not in interacting field $A_{\mu}$
(as it actually takes place in Refs.\,\cite{efr, bla2}), but in free field
$A_{\mu}^{(0)}$. The coefficient functions of this expansion, called effective
amplitudes, represent rather complicated combinations of HTL-amplitudes and
an equilibrium propagator for a soft gluon. They preserve many features of
usual HTL-amplitudes, but have more direct physical meaning -- these functions
set on mass-shell plasma excitations representing the scattering amplitudes of
soft collective modes.

At present there are a few methods of constraction of relativistic kinetic
equations (see, e.g. review in Ref.\,\cite{mar3}). A more powerful and more
convenient tool to derive relativistic Boltzmann equations from exact field
Scwinger-Dyson equations, the so-called closed-time-path (CTP) formalism of
nonequilibrium quantum-field theory, is considered. This general method is
suitable to the same extent for deriving transport equations both hard modes
and soft ones of plasma medium. From most papers close to the
subject of our research, where for derivation of the Boltzmann equations for
soft gluon longitudinal and transverse quasiparticles the CTP formalism is
used, it should be mentioned the work by Ni\'egawa \cite{nie}. Our view on
the problem of derivation of the Boltzmann equations for soft plasma modes
is quite different from a view presented in Ref.\,\cite{nie}. Since
we are interested only with the processes of nonlinear interaction of the
soft excitations among themselves and with hard termal particles,
the use of quasiclassical approximation is justified. By virtue of this fact
we consider that it has no need to use tools of nonequilibrium field theory
which are general, but for our particular problems, are needlessly complicated.

Our approach is based on the fundamental system of dynamical equations derived
by Blaizot and Iancu \cite{bla2}, which is a local formulation of the
HTL-equations of motion for soft fluctuating gluon and quark fields and their
induced sources. These equations are obtained by means of a truncation of the
Schwinger-Dyson hierarchy and already contain the maximum
comprehensive information at leading order in $g$ about dynamics of hot QCD
plasma at soft momentum scale\footnote{However it should be noted that the
approximation used here may be unsuitable in attempt of calculation of
effective amplitude of elastic plasmon-plasmon collision in the limit
of vanishing momentum transfer,
near to "forward scattering".}. Therefore this system is the most
convenient and optimal start point for derivation of transport equations for
soft
plasma excitations. In our previous paper \cite{mar3} it was shown how starting
from similar dynamic equations one can derive collision term for elastic
scattering of two colorless plasmons. However, as was mentioned above,
the direct approach presented there, is very cumbersome and complicated for its
extension to the processes of collisions with account of many colorless
plasmons. For this reason in the present paper and accompanying ones,
proceed from quasiclassical character of a problem, we initially assume
that a structure of the collision terms are determined by Fermi's golden rule
and thus we focus our efforts on calculation exclusively of the scattering probabilities,
which in fact determine all dynamics of interaction of soft boson excitations.

The following step is to extend analysis  carried out in these papers
on the fermion degree of freedom of plasma excitations in hot QCD plasma.
Here, the series of new specific moments, which will be fully considered
in the subsequent works, is arizen.

The paper I is organized as follows. In section 2 preliminary comments,
with regard to derivation of the Boltzmann equation, describing
$(2n + 2)$-plasmon decay processes, are given. In section 3 all the
conventions and the notations used in this paper, are summarized, and the
nonlinear integral equation for gauge potential $A_{\mu}$, playing key
position in our subsequent research, is written out. Section 4 is devoted
to explanation of the correspondence principal that enables one to establish
a direct connection of the matrix elements of elastic collision of
$n + 1$ colorless plasmons with certain effective current, generating
these processes. Section 5 presents a detailed consideration of the
Boltzmann equation for four-plasmon decay. In section 6 we estimate the
typical value  of plasmon occupation numbers, wherein one can rectricted
the consideration to taking into account only contribution from four-plasmon
decay or it should be considered all higher decay processes. In section 7
a complete algorithm of succesive calculation of the matrix elements
defining $(2n + 2)$-plasmon decay processes in the temporal gauge, is presented.
In section 8 we discuss a problem of a gauge independence of these matrix
elements. In section 9 we give some speculation  concerning extension
of an approach proposed in this work to research of dynamics of colorless
plasmons to the case of color plasmons. In closing section, when the main
results were outlined, we briefly discuss the additional difficulties
associated with appearance of the nonlinear frequency shift of longitudinal
oscillations generated by plasmon collective interactions.\\

\section{Preliminares}
\setcounter{equation}{0}

The main part of our work will be concerned with the constraction of an
effecive kinetic theory for soft colorless longitudinal excitations,
propagating in a purely gluonic plasma, with no
quarks. In other words we assume that a localized number density
of plasmons $N^l ({\bf p}, x)\equiv(N_{\bf p}^{l\,ab})$ is diagonal
in color space
\[
N_{\bf p}^{l \, ab} = \delta^{ab} N_{\bf p}^l ,
\]
where $a,b = 1, \ldots , N_c^2 - 1$ for $SU(N_c)$ gauge group, and we
consider the change of the number density of the colorless plasmons
$N_{\bf p}^l$ as a result of their interactions among themselves.
In section 9 only using some general reasonable assumptions,
we made an attempt of an extension of the obtained results to the case of
color excitations with non-diagonal plasmon number density.

The dispersion relation for plasmons $\omega^l = \omega^l ({\bf p}) \equiv
\omega_{\bf p}^l$ is defined by
\begin{equation}
{\rm Re} \,^{\ast}\!\Delta^{\!-1\,l} (\omega, {\bf p}) = 0 ,
\label{eq:2q}
\end{equation}
where
\[
\,^{\ast}\!\Delta^{\!-1\,l}(\omega, {\bf p}) = p^2\bigg(
1 + \frac{3 \omega_{pl}^2}{{\bf p}^2}
\bigg[1 - F \bigg( \frac{\omega}{\vert {\bf p} \vert} \bigg) \bigg]\bigg)\;,\;
F (x) \equiv \frac{x}{2} \bigg[ \ln \bigg \vert \frac{1 + x}{1 - x}
\bigg \vert - i \pi
\theta ( 1 - \vert x \vert ) \bigg]
\]
is an inverse resummed longitudinal propagator and
$\omega_{pl}^2 = g^2 N_cT^2/9$ is a plasma frequency square.
From physical considerations it is clear that the notions of the plasmon
and accordingly, plasmon number density have a meaning only with use of the
condition
\[
\omega^l \tau^l \gg 1,
\]
i.e. for the plasmon frequency that is much more than inverse lifetimes of
plasmon $1/\tau^l$. Generally speaking, this time is defined by just as the
processes of plasmons scattering off each other, so by the scattering processes
including soft modes and hard termal particles (hard thermal gluons in our case).
For sufficiently large intensity of plasma excitations (exact estimations will
be given in Sec.\,6 of this paper and in \cite{mar2}), the processes of a first
type can play the same important role in dynamics of a system as the second ones
and even become dominant. Under this condition, plasma may be considered as
involving two interacting subsystems: the subsystem of hard thermal gluons and
the subsystem of soft plasmons, which exchange energy among themselves. As
was mentioned in Introduction, here, we have restricted ourselves to
detailed consideration of the processes of interactions in the plasmon
subsystem. This fact is embodied in the structure of collision term in the
plasmon kinetic equation proposed below.

We expect the time-space evolution of scalar function $N_{\bf p}^l$ to be
described by
\begin{equation}
\frac{\partial N_{\bf p}^l}{\partial t} +
{\bf V}_{\bf p}^l\cdot\frac{\partial N_{\bf p}^l}{\partial {\bf x}} =
- N_{\bf p}^l \Gamma_{\rm d} [ N_{\bf p}^l ] + ( 1 + N_{\bf p}^l )
\Gamma_{\rm i}[ N_{\bf p}^l ] ,
\label{eq:2w}
\end{equation}
where ${\bf V}_{\bf p}^l = \partial \omega_{\bf p}^l / \partial {\bf p}$ is a
group velocity of the longitudinal oscillations. The generalized decay rate
$\Gamma_{\rm d}$ and inverse decay rate $\Gamma_i$ in general case are
(non-linear) functionals dependent on the plasmon number density.
From here on such a functional dependence is denoted by argument of
a function in square brackets.

We shall consider that the decay rate $\Gamma_{\rm d}$ and regenerating
rate $\Gamma_{\rm i}$ can be formally represented in the form of functional
expansion in powers of the plasmon number density
\begin{equation}
\Gamma_{\rm d} [N_{\bf p}^l] = \sum_{n = 1}^{\infty}
\Gamma_{\rm d}^{(2n + 1)} [N_{\bf p}^l] , \; \;
\Gamma_{\rm i} [N_{\bf p}^l] = \sum_{n = 1}^{\infty}
\Gamma_{\rm i}^{(2n + 1)} [N_{\bf p}^l],
\label{eq:2e}
\end{equation}
where
\begin{equation}
\Gamma_{\rm d}^{(2n + 1)} [N_{\bf p}^l] =
\int\!d{\cal T}^{(2n + 1)} \, {\it w}_{2n + 2}
({\bf p}, {\bf p}_1, \ldots ,{\bf p}_n;
{\bf p}_{n + 1}, \ldots , {\bf p}_{2n + 1})
\, N_{{\bf p}_1}^l \ldots N_{{\bf p}_n}^l
\label{eq:2r}
\end{equation}
\[
\times(1 + N_{{\bf p}_{n + 1}}^l) \ldots
(1 + N_{{\bf p}_{2n + 1}}^l) ,
\]
\begin{equation}
\Gamma_{\rm i}^{(2n + 1)} [N_{\bf p}^l] =
\int\!d{\cal T}^{(2n + 1)} \,
{\it w}_{2n + 2}({\bf p}, {\bf p}_1, \ldots ,{\bf p}_n;
{\bf p}_{n + 1}, \ldots , {\bf p}_{2n + 1})
\,(1 +  N_{{\bf p}_1}^l) \ldots (1 + N_{{\bf p}_n}^l)
\label{eq:2t}
\end{equation}
\[
\times N_{{\bf p}_{n + 1}}^l \ldots N_{{\bf p}_{2n + 1}}^l .
\]
Here, ${\it w}_{2n + 2} \equiv {\it w}_{2n + 2}({\bf p}, {\bf p}_1,
\ldots ,{\bf p}_n; {\bf p}_{n + 1}, \ldots , {\bf p}_{2n + 1})$
is a scattering probability for process of elastic collision of
$n + 1$ plasmons, and phase-space measure is
\begin{equation}
\int\! d{\cal T}^{(2n + 1)}\!\equiv\!
\int (2 \pi)^4 \delta^{(3)}({\bf p}_{\rm in} - {\bf p}_{\rm out})
\delta(E_{\rm in} - E_{\rm out})\!
\prod_{k =1}^{2n+1}\!\frac{d{\bf p}_{k}}{(2\pi)^3},
\label{eq:2y}
\end{equation}
where
\[
{\bf p}_{\rm in}={\bf p}+{\bf p}_1+\ldots +{\bf p}_n, \quad
E_{\rm in} = \omega^l_{\bf p}+\omega^l_{{\bf p}_1} + \ldots
+\omega^l_{{\bf p}_n},
\]
and
\[
{\bf p}_{\rm out}={\bf p}_{n+1} + \ldots + {\bf p}_{2n+1}, \quad
E_{\rm out} = \omega^l_{{\bf p}_{n+1}} +\ldots +\omega^l_{{\bf p}_{2n+1}}.
\]
The Dirac $\delta$-functions in Eq.\,(\ref{eq:2y}) expresses the momentum
and  energy
conservation in the collision process. In writing the expressions
(\ref{eq:2r}) and (\ref{eq:2t}), we have used the fact that the probabilities
of direct and reverse processes are identical.

We make some comments concerning Eqs.\,(\ref{eq:2e})\,--\,(\ref{eq:2t}).
By virtue of the fact that
processes of nonlinear interaction of odd number of the plasmons are
kinematically forbidden by the conservation laws, in expansions of generalized
rates (\ref{eq:2e}) we leave only odd terms in powers $N_{\bf p}^l$.
Furthermore a structure of the terms of the expansions (\ref{eq:2r})
and ({\ref{eq:2t})
is chosen such that it taken into account the scatåering processes only with
equal number of the plasmons prior to interaction and upon it, i.e. the
scattering processes of the following type\footnote{The term {\it the decay
processes} will be also used for the processes of (\ref{eq:2u}) type.}
\[
{\rm g}^{\ast} + {\rm g}_1^{\ast} \rightleftharpoons
{\rm g}_2^{\ast} + {\rm g}_3^{\ast},\quad {\rm for}\;n =1,
\]
\begin{equation}
{\rm g}^{\ast} + {\rm g}_1^{\ast} + {\rm g}_2^{\ast} \rightleftharpoons
{\rm g}_3^{\ast} + {\rm g}_4^{\ast} + {\rm g}_5^{\ast},\quad {\rm for}\;n = 2,
\label{eq:2u}
\end{equation}
\[
\ldots,
\]
where ${\rm g}^{\ast}, \, {\rm g}_1^{\ast}, \ldots$ are the plasmon collective
excitations. Such a choose of the structure of the decay and regenerating
rates assumes that decay processes with unequal even number of incoming
and outgoing soft external legs of the type
\[
{\rm g}^{\ast} + {\rm g}_1^{\ast}  \rightleftharpoons
{\rm g}_2^{\ast} + {\rm g}_3^{\ast} + {\rm g}_4^{\ast}
+ {\rm g}_5^{\ast},\quad{\rm for}\;n = 2,
\]
etc, are kinematically suppressed as compared with ``elastic'' scattering
processes (\ref{eq:2u}) (speaking  more exactly,
kinematic regions of momentum variables in ``elastic'' and ``inelastic''
scattering accessible by conservation laws, are not be covered each
other, and a contribution of the last processes
to the nonlinear plasmon dynamics to the order of interest is not important).

The scattering probability ${\it w}_{2n + 2}$ must satisfy the symmetry relations
over permutations of external momenta
\begin{equation}
{\it w}_{2n + 2}({\bf p}, {\bf p}_1, \ldots ,{\bf p}_n;
{\bf p}_{n + 1}, \ldots , {\bf p}_{2n + 1}) =
{\it w}_{2n + 2}({\bf p}_1, {\bf p}, \ldots ,{\bf p}_n;
{\bf p}_{n + 1}, \ldots , {\bf p}_{2n + 1}) = \ldots
\label{eq:2i}
\end{equation}
\[
={\it w}_{2n + 2}({\bf p}_n, {\bf p}_1, \ldots ,{\bf p};
{\bf p}_{n + 1}, \ldots , {\bf p}_{2n + 1}) =
{\it w}_{2n + 2}({\bf p}_{n + 1}, \ldots ,{\bf p}_{2n + 1};
{\bf p}, {\bf p}_1, \ldots , {\bf p}_n),
\]
which are a consequence of an indistinguishability of the colorless
plasmons. The special consequences of symmetry properties (\ref{eq:2i}) are
the conservation laws of energy momentum and total number of colorless
plasmons in the decay process (\ref{eq:2u})
\[
{\cal E} \equiv \int\!\frac{d{\bf p}}{(2 \pi)^3} \;
\omega^l_{\bf p} N^l_{\bf p} =  const, \; \;
{\bf {\cal K}}
\equiv \int\!\frac{d{\bf p}}{(2 \pi)^3} \;
{\bf p} \, N^l_{\bf p} =  const,\;\;
\]
\[
{\cal N} \equiv \int\!\frac{d{\bf p}}{(2 \pi)^3} \;
N^l_{\bf p} = const.
\]\\

\section{\bf  Blaizot-Iancu equations. Correlation function of the
boson excitations}
\setcounter{equation}{0}

We adopt conventions of Blaizot and Iancu \cite{bla2}.
We use the metric $g^{\mu \nu} = diag(1,-1,-1,-1)$, choose units such
that $c=k_{B}=1$ and note $X=(X_0,{\bf X}), \,p=(p_0\equiv\omega,{\bf p}).$
The gauge field $A_{\mu}=A_{\mu}^{a}t^{a}$ with
$N_{c}^{2}-1$ Hermitian generators in the fundamental representation obeys
the field equation
\begin{equation}
[D^{\nu},F_{\mu \nu}(X)] -
\xi_0^{-1} n_\mu n^\nu A_{\nu}(X) = j_{\mu}(X),
\label{eq:3q}
\end{equation}
where $D_{\mu} = \partial_{\mu} + igA_{\mu}(X)$ is a covariant derivative,
$F_{\mu \nu}=F_{\mu \nu}^{a}t^{a}$ is a field strength tensor with
$F_{\mu \nu}^{a} = \partial_\mu A_{\nu}^{a} - \partial_\nu A_{\mu}^{a}-
gf^{abc}A_{\mu}^{b}A_{\nu}^{c}$; $[\,,\,]$ denotes a commutator and
$ \xi_0$ is a gauge parameter fixing a temporal gauge. In what follows
4-vector $n_\mu$ will be identified with  global 4-velocity $u_\mu$
of the plasma.

The color current $j_{\mu}$ on the right-hand side of field
equation (\ref{eq:3q}) can be written as
\begin{equation}
j_{\mu}(X) = 2gN_c\!\int\!\!\frac{d{\bf k}}{(2 \pi)^3} \, v_{\mu}
\delta N({\bf k},X),\quad v=(1,{\bf v}),\;{\bf v}={\bf k}/\vert{\bf k}\vert,
\label{eq:3w}
\end{equation}
where $\delta N({\bf k},X) = \delta N^{a}t^a$
is a soft fluctuation in the gluon color density.
We suppose that there is no external color current and/or mean
color field in the system, and therefore an expectation value of the
induced color current (\ref{eq:3w}) over the off-equilibrium ensemble
equals zero, i.e. $\langle j_{\mu}(X)\rangle = 0$. The consequence of this
fact is $\langle\delta N({\bf k},X)\rangle = 0$.

On a space-time scale $(gT)^{-1}$ the soft fluctuation $\delta N({\bf k},X)$
satisfies the following equation:
\begin{equation}
[v \cdot D_X, \delta N({\bf k}, X)] =
i g\langle[v \cdot A(X), \delta N({\bf k}, X)]\rangle
-\,g\{\,{\bf v}\cdot {\bf E}(X) - \langle {\bf v}\cdot {\bf E}(X)\rangle\}
\, \frac{{\rm d} N (\epsilon_k)}{{\rm d} \epsilon_k} .
\label{eq:3e}
\end{equation}
Here, ${\bf E} (X) = {\bf E}^a(X) t^a$ is a chromoelectric field,
$E^i = F^{i0}$; $N(\epsilon_k) = 1/(\exp(\epsilon_k/T) - 1)$ is a boson
occupation factor and $\epsilon_k \equiv\vert{\bf k}\vert$.
On the right-hand side of Eq.\,(\ref{eq:3e}) we have entered the additional
average terms,
so that Eq.\,(\ref{eq:3e}) will be equal zero for the expectation value taken
over the off-eqilibrium ensembles. The self-consistent system of
Eqs.\,(\ref{eq:3q})\,--\,(\ref{eq:3e}) for a soft fluctuating field $A_\mu$
and their induced current presents a pure gauge sector of general system
of self-consistent equations
first obtained by Blaizot and Iancu in Ref.\,\cite{bla2}.

The equations (\ref{eq:3q})\,--\,(\ref{eq:3e}) are solved by the approximation
scheme method -- {\it the weak-field expansion}. For this purpose first of all
we expand the soft fluctuation of the gluon color density
\begin{equation}
\delta N = \sum_{s=1}^{\infty} \delta N^{(s)},
\label{eq:3r}
\end{equation}
where index $s$ shows that $\delta N^{(s)}$
is proportional to the $s$th power of $A_{\mu}$.
The expansion of a color current, corresponding to (\ref{eq:3r}) has the form
\begin{equation}
j_{\mu} = \sum_{s=1}^{\infty} j^{(s)}_{\mu},\quad
j_{\mu}^{(s)} = 2gN_c\!\int\!\frac{d{\bf k}}{(2 \pi)^3} \, v_{\mu}
\delta N^{(s)}.
\label{eq:3t}
\end{equation}

Now we turn to the field equation (\ref{eq:3q}),
connecting the gluon soft field with induced color current $j_{\mu}$.
Let us rewrite this equation, explicitly separating free parts
in (\ref{eq:3q}) from interacting terms. We have
\begin{equation}
\partial^{\nu}(F_{\mu \nu})_{L} - \xi_0^{-1} n_{\mu}
n^{\nu}A_{\nu} - j^{(1)}_{\mu} =
j_{NL\,{\mu}} - ig\,\{\partial^{\nu}[A_{\mu},A_{\nu}] -
\langle\partial^{\nu}[A_{\mu},A_{\nu}]\rangle\} -
\label{eq:3y}
\end{equation}
\[
-\,ig\{[A^{\nu},(F_{\mu \nu})_{L}] - \langle[A^{\nu},(F_{\mu \nu})_{L}]\rangle\}
+\,g^{2}\{[A^{\nu},[A_{\mu},A_{\nu}]] -
\langle [A^{\nu},[A_{\mu},A_{\nu}]]\rangle\}.
\]
Here, indices $L$ and $NL$ denote linear and nonlinear parts of
strength  tensor and the color induced current with respect to $A_{\mu}$.
As in the case of dynamical equation (\ref{eq:3e}), we enter the additional
averaged terms on the right-hand side of Eq.\,(\ref{eq:3y}), which ensure vanishing
left and right parts when an expectation value is taken.

Substituting the expansion (\ref{eq:3r}) into (\ref{eq:3e}) and collecting
the terms of the same order in $A_{\mu}$, we derive the system of equations for
$\delta N^{(s)}, s=1, 2, \ldots,$ whose solution is easily obtained by
Fourier transformation. Substituting such a derived expression we obtain
the explicit form of the terms in a color current expansion (\ref{eq:3t})
\begin{equation}
j_{\mu}^{(s)} (X) = t^a\!\!\int\!dp \, j^{(s)a}_{\mu} (p) \,{\rm e}^{-i p\cdot X},
s=1, 2, \ldots,
\label{eq:3u}
\end{equation}
where
\[
j^{(1)a}_{\mu}(p)= \Pi_{\mu \nu}(p)A^{a\nu}(p),
\]
\[
j^{(2)a}_{\mu}(p)= \frac{1}{2!}\,g\!\int\!
\delta\Gamma^{a a_1 a_2}_{\mu \mu_1 \mu_2}(p,-p_{1},-p_{2})
\{A^{a_1\mu_1}(p_{1})A^{a_2\mu_2}(p_{2})
-\,\langle A^{a_1\mu_1}(p_{1})A^{a_2\mu_2}(p_{2})\rangle\}
\]
\[
\times\delta (p - p_{1} - p_{2}) dp_{1}dp_{2},
\]
\[
\ldots,
\]
\[
j^{(s)a}_{\mu}(p)= \frac{1}{s!}\,g^{s-1}\!\int
\!\delta\Gamma^{a a_1\ldots a_s}_{\mu\mu_1\ldots\mu_s}(p,-p_{1},\ldots,-p_{s})
\{A^{a_1\mu_1}(p_{1})A^{a_2\mu_2}(p_{2})\ldots A^{a_s\mu_s}(p_{s})-
\]
\[
- A^{a_1\mu_1}(p_{1})\langle A^{a_2\mu_2}(p_{2})\ldots
A^{a_s\mu_s}(p_{s})\rangle -\,\ldots\, -
\langle A^{a_1\mu_1}(p_{1})A^{a_2\mu_2}(p_{2})\ldots
A^{a_s\mu_s}(p_{s})\rangle\}
\]
\[
\times\delta (p - \sum_{i=1}^{s}p_{i})\prod_{i=1}^{s}dp_{i},\;\ldots\,.
\]
Here, $\Pi_{\mu\nu}(p)$ is a soft-gluon self-energy and
$\delta\Gamma^{a a_1\ldots a_s}_{\mu\mu_1\ldots\mu_s}$ are usual HTL-functions
\cite{bra1, bla2}. Rewriting equation (\ref{eq:3y}) in momentum space and taking
into account (\ref{eq:3u}) we lead to the nonlinear integral equation for
gauge potential $A_{\mu}$, playing key position in our subsequent research
\begin{equation}
\,^{\ast}\tilde{\cal D}^{-1 \, \mu \nu}(p) A^{a}_{\nu}(p) =
-J^{a\mu}_{NL}[A]\equiv-\sum_{s=2}^{\infty} J^{(s)a\mu}(A,\ldots,A),
\label{eq:3i}
\end{equation}
where members of a series on the right-hand side of Eq.\,(\ref{eq:3i}) have
a structure
\begin{equation}
J^{(s)a}_{\mu}(A,\ldots,A) =  \frac{1}{s!}\,g^{s-1}\!\int\!
\,^{\ast}\Gamma^{a a_1\ldots a_s}_{\mu\mu_1\ldots\mu_s}(p,-p_{1},\ldots,-p_{s})
\{A^{a_1\mu_1}(p_{1})A^{a_2\mu_2}(p_{2})\ldots A^{a_s\mu_s}(p_{s})
\label{eq:3o}
\end{equation}
\[
- A^{a_1\mu_1}(p_{1})\langle A^{a_2\mu_2}(p_{2})\ldots
A^{a_s\mu_s}(p_{s})\rangle -\,\ldots\, -
\langle A^{a_1\mu_1}(p_{1})A^{a_2\mu_2}(p_{2})\ldots
A^{a_s\mu_s}(p_{s})\rangle\}
\]
\[
\times\delta (p - \sum_{i=1}^{s}p_{i})\prod_{i=1}^{s}dp_{i}.
\]
Here, the coefficient functions
$\,^{\ast}\Gamma^{a a_1\ldots a_s}_{\mu\mu_1\ldots\mu_s}$ for $s=3$
and $s=4$ represent a sum of a bare vertices and corresponding HTL-corrections,
and for $s > 4$ we rename $\,^{\ast}\Gamma^{(s)}\equiv\delta\Gamma^{(s)}$.
$\,^{\ast}\tilde{\cal D}^{\mu \nu}(p)$ is a medium modified
(retarded) gluon propagator in a temporal gauge
\begin{equation}
\,^{\ast}\tilde{\cal D}_{\mu \nu}(p)= -
P_{\mu \nu}(p) \,^{\ast}\!\Delta^t(p) -
\tilde{Q}_{\mu \nu}(p) \,^{\ast}\!\Delta^l(p) +
\xi_0\frac{p^2}{(p\cdot u)^2} D_{\mu \nu}(p),
\label{eq:3p}
\end{equation}
where $\,^{\ast}\!\Delta^{t,l}(p) = 1/(p^2 - \Pi^{t,l}(p)), \,
\Pi^t(p) = \frac{1}{2} \Pi^{\mu \nu}(p) P_{\mu \nu}(p), \,
\Pi^l(p) = \Pi^{\mu \nu}(p)\tilde{Q}_{\mu \nu}(p)$.
The Lorentz matrices in (\ref{eq:3p}) are defined by
\begin{equation}
P_{\mu \nu} (p) = g_{\mu \nu} - u_{\mu}u_{\nu} - \frac{(p\cdot u)^2}{p^2}
\tilde{Q}_{\mu \nu}(p), \quad
\tilde{Q}_{\mu \nu}(p) =
\frac{\tilde{u}_{\mu}(p)\tilde{u}_{\nu} (p)}{\bar{u}^2(p)},\quad
D_{\mu \nu}(p)=\frac{p_{\mu}p_{\nu}}{p^2},
\label{eq:3a}
\end{equation}
\begin{equation}
\tilde{u}_{\mu}(p) = \frac{p^2}{(p\cdot u)}(p_{\mu} - u_{\mu} (p\cdot u)),
\quad \bar{u}=p^2u_{\mu} - p_{\mu}(p\cdot u).
\label{eq:3s}
\end{equation}
Let us assume that we are in a rest frame of a heat bath, so that
$u_{\mu}=(1,0,0,0).$

At the end of this section we would like to
introduce the correlation function of the soft-bosonic
excitations,
\begin{equation}
I_{\mu \nu}^{ab}(p^{\prime},p)= \langle A_{\mu}^{\ast\,a}(p^{\prime})
A_{\nu}^{b}(p) \rangle.
\label{eq:3d}
\end{equation}
The asterisk denotes the complex conjugate. The considered soft-gluon
excitations are necessarily {\it colorless} by virtue of the fact that
mean field $\langle A_{\mu}^a \rangle$ or associated mean induced color
current are assumed to be vanishing. Therefore, for the physical situation of
interest, the off-equilibrium two-point function (\ref{eq:3d}) is diagonal
in a color space.

For the conditions of stationary and homogeneous of hot gluon plasma we have
\begin{equation}
I_{\mu \nu}^{ab}(p^{\prime},p)=\delta^{ab}I_{\mu \nu}(p^{\prime})
\delta (p^{\prime}-p).
\label{eq:3f}
\end{equation}
The off-equilibrium perturbations which are slowly varying in space and
time lead to a $\delta$-function broadering, and $I_{\mu\nu}(p^{\prime},p)$
depends on both arguments $p$ and $p^{\prime}$. Let us introduce
$I_{\mu \nu}(p^{\prime},p)=I_{\mu \nu}
(p, \triangle p)$, $\triangle p = p^{\prime} - p$ with
$\vert\,\triangle p / p\,\vert\ll\!1$,
and insert the correlation function in the Wigner form
\begin{equation}
I_{\mu \nu}(p,x)= \int I_{\mu \nu}(p, \triangle p)
\,{\rm e}^{- i \triangle p\cdot x} d \triangle p,
\label{eq:3g}
\end{equation}
slowly depending on $x=(t,{\bf x})$. In global equilibrium hot QCD plasma the oscillations
of two types: the longitudinal and transverse ones can be extended. In this
connection we define the Wigner function $I_{\mu \nu}(p,x)$ in the form
of an expansion
\begin{equation}
I_{\mu \nu}(p,x)= P_{\mu\nu}(p)I_{p}^t + \tilde{Q}_{\mu\nu}I_{p}^l,\quad
I_{p}^{(t,l)}\equiv I^{(t,l)}(p,x).
\label{eq:3h}
\end{equation}\\

\section{\bf The correspondence principle}
\setcounter{equation}{0}

The main purpose of this section is a derivation in an explicit form
of the scattering probability ${\it w}_{2n+2}$ for process of elastic
collision of $n+1$ plasmons in hot gluon plasma.
In our previous paper \cite{mar3} we have derived the probability ${\it w}_4$
for a two-to-two scattering process of colorless plasmons.
It was calculated by simple extracting all possible contributions to this process.
However, use of such a direct approach in a general case of the process
of elastic scattering of $n+1$ plasmons becomes ineffective because of
awkwardness and complicated computations. In this case the method
developed by Tsytovich \cite{gai, tsy} in the theory of nonlinear processes in
electron-ion plasma, known as the correspondence principle, is more
convenient for deriving explicit form of the scattering probability
${\it w}_{2n + 2}$. We have already mentioned it in Ref.\,\cite{mar3}.
Here, this approach will be
developed as applied to general case of $(2n+2)$-plasmon decay. For non-Abelian
plasma this approach is especially effective in the temporal gauge, when we have
a closer correspondence with the electrodynamics of an ordinary plasma.
The gist of this method is as follows.

The change in the colorless plasmon numbers, caused by spontaneous processes of
plasmon decays only, is
\[
\left( \frac{\partial N_{\bf k}^{l}}{\partial t} +
{\bf V}_{\bf p}^{l}\cdot\frac{\partial N_{\bf p}^{l}}
{\partial {\bf x}} \right)^{sp}\! =\!
\sum_{n=1}^{\infty}
\int\! d{\cal T}^{(2n + 1)} \, {\it w}_{2n + 2}
({\bf p}, {\bf p}_1, \ldots ,{\bf p}_n;
{\bf p}_{n + 1}, \ldots , {\bf p}_{2n + 1})
\, N_{{\bf p}_1}^l \ldots N_{{\bf p}_{2n + 1}}^l.
\]
This equation follows from (\ref{eq:2w}) in the limit of a small
intensity $N_{\bf p}^l \rightarrow 0$ and using the fact that
the occupation numbers $N_{{\bf p}_i}^l$ are larger than one,
$N_{{\bf p}_i}^l +1 \approx N_{{\bf p}_i}^l$. In this case the change of
energy of the longitudinal excitations is
\begin{equation}
\left( \frac{{\rm d}{\cal E}}{{\rm d} t} \right)^{sp} =
\frac{\rm d}{{\rm d}t}\left(
\int\!\frac{d{\bf p}}{(2 \pi)^3} \;
\omega^l_{\bf p} N^l_{\bf p}\right)
\label{eq:4q}
\end{equation}
\[
=\sum_{n=1}^{\infty}
\int\!\frac{d{\bf p}}{(2 \pi)^3}
\int\!d{\cal T}^{(2n + 1)}\,\omega_{\bf p}^l{\it w}_{2n + 2}
({\bf p}, {\bf p}_1, \ldots ,{\bf p}_n;
{\bf p}_{n + 1}, \ldots , {\bf p}_{2n + 1})
N_{{\bf p}_1}^l \ldots N_{{\bf p}_{2n + 1}}^l.
\]
On the other hand the value $({\rm d}{\cal E}/{\rm d}t)^{sp}$ represents the
emitted radiant power of the longitudinal waves ${\cal I}^l$, which in turn is
equal to the work done by the radiation field with the color current,
creating it, per unit time
\begin{equation}
{\cal I}^l = \lim\limits_{\tau,\,V\rightarrow\infty}
\frac{1}{\tau V}\!\int\limits_{-\tau/2}^{\tau/2}\!\int\limits_{V}
d{\bf x}dt\,\langle {\bf E}^a ({\bf x},t)\cdot
{\bf J}^a ({\bf x},t) \rangle =
\lim\limits_{\tau,\,V\rightarrow\infty}
\frac{(2\pi)^4}{\tau V}
\int\!d{\bf p}d\omega\,\langle {\bf E}^a ({\bf p},\omega)\cdot
{\bf J}^a ({\bf p},\omega) \rangle.
\label{eq:4w}
\end{equation}
Here, $V$ is a spatial volume of integration,
$E^{ai}({\bf x},t) = -\,\partial A^{ai}({\bf x},t)/ \partial t$
is a chromoelectric field in the temporal gauge,
$E^{ai} ({\bf p},\omega)$ and $J^{ai} ({\bf p},\omega)$ are the Fourier
components of field and current, correspondingly.
Averaging over the time was made for
elimination of oscillating terms in ${\cal I}^l$. For a transformation
(\ref{eq:4w}) we use a relation
\[
\lim\limits_{\tau\rightarrow\infty}\!\int\limits_{-\tau/2}^{\tau/2}
\!{\rm e}^{i\omega t}
dt=2\pi\delta(\omega).
\]
The Fourier-component of a field
${\bf E}^a({\bf p}, \omega) = {\bf p} E^a({\bf p},\omega)/\vert {\bf p}\vert$
is associated with ${\bf J}^a({\bf p}, \omega)$ by the field equation
\begin{equation}
E^a({\bf p}, \omega) = i\,\frac{p^2}{\omega}
\,^{\ast}\!\Delta^{l}( {\bf p},\omega)
\,\frac{({\bf p} \cdot {\bf J}^a({\bf p}, \omega))}
{\vert {\bf p} \vert} .
\label{eq:4e}
\end{equation}
For weak-absorption medium, when
${\rm Im}\,^{\ast}\!\Delta^{\!-1\,l}( {\bf p},\omega)\rightarrow 0$,
the propagator $\,^{\ast}\!\Delta^{l}( {\bf p},\omega)$ can be approximated in
the following way
\[
\,^{\ast}\!\Delta^{l}(p)=\frac{1}{p^2-\Pi^l(p)}
\approx\frac{{\rm P}}{{\rm Re}\,(p^2 - \Pi^l(p))}
-i\pi {\rm sign}(\omega)\delta({\rm Re}\,^{\ast}\!\Delta^{\!-1\,l}(p)).
\]
Here, we consider that
${\rm sign}({\rm Im}\,^{\ast}\!\Delta^{\!-1\,l}(p))\!=\!{\rm sign}(\omega)$.
The symbol
${\rm P}$
denotes a principal value. Substituting Eq.\,(\ref{eq:4e}) into (\ref{eq:4w})
and taking into account reality of ${\cal I}^l$, we derive
\begin{equation}
{\cal I}^l = -\pi\!\lim\limits_{\tau,\,V\rightarrow\infty}
\frac{(2\pi)^4}{\tau V}\!
\int\!dp\,\omega\,{\rm sign}(\omega)\tilde{Q}^{\mu\mu^{\prime}}(p)
\,\langle J^{\ast a}_{\mu}(p)J^a_{\mu^{\prime}}(p)\rangle
\delta({\rm Re}\,^{\ast}\!\Delta^{\!-1\,l}(p)).
\label{eq:4r}
\end{equation}
In derivation of the last expression, we present a longitudinal projector
${\bf p}\otimes{\bf p}/{\bf p}^2$ in a Lorentz covariant form
(Eqs.\,(\ref{eq:3a}) and (\ref{eq:3s}))
\begin{equation}
\frac{{\bf p}\otimes{\bf p}}{{\bf p}^2}\rightarrow -\,\frac{\omega^2}{p^2}
\,\tilde{Q}^{\mu\mu^{\prime}}(p).
\label{eq:4t}
\end{equation}
Since the nonlinear part of a current only is responsible for processes of
plasmon decays, then it is necessary to set $J^{a\mu} = J^{a\mu}_{NL}$ in
the expression (\ref{eq:4r}), where $J^{a\mu}_{NL}$
is defined by equation (\ref{eq:3i}). We perceive
the $\delta$-function of the real part of inverse longitudinal propagator
in the ordinary sense
\begin{equation}
\delta({\rm Re}\,^{\ast}\!\Delta^{\!-1\,l}({\bf p},\omega))=
\frac{1}{2\omega_{\bf p}^l}\,{\rm Z}_l({\bf p})\,[
\delta(\omega-\omega_{\bf p}^l) +
\delta(\omega+\omega_{\bf p}^l)],
\label{eq:4y}
\end{equation}
where ${\rm Z}_l({\bf p})$ is the residue of the effective gluon propagator
at the plasmon pole.

Substituting the nonlinear current expansion (\ref{eq:3i}) into a correlation
function in integrand of Eq.\,(\ref{eq:4r}) we face the product of two
series. However in this product it is necessary to leave only a sum of a product
of terms having the same order in power of a potential $A_{\mu}^a$. Then
we have equal number of complex-conjugate potentials and the non-conjugate
ones between the inside of the angular brackets of statistical averaging.
The necessity of such a choose of terms was demonstrated by a direct
calculation of the Boltzmann equation for four-plasmon decay in
Ref.\,\cite{mar3}.
In this case only a desired $\delta$-functions entering to
integration measure $\int\!d{\cal T}^{(3)}$, expressing the energy
and the momentum conservation of the four-plasmon decay, arise. We assume this
for general rule, which is valid for decay process with arbitrary number of
plasmons. Thus the emitted radiant power (\ref{eq:4r}), taking into account
emission caused by plasmon decay processes only, can be presented in
the form of expansion
\begin{equation}
{\cal I}^l=\sum_{s=2}^{\infty}{\cal I}^{l\,(s)},
\label{eq:4u}
\end{equation}
where
\begin{equation}
{\cal I}^{l\,(s)} = -\pi\!\lim\limits_{\tau,\,V\rightarrow\infty}
\frac{(2\pi)^4}{\tau V}\!
\int\!dp\,\omega\,{\rm sign}(\omega)\,
\tilde{Q}_{\mu\mu^{\prime}}(p)
\,\langle J^{\ast (s)a\mu}_{NL}(p)J^{(s)a\mu^{\prime}}_{NL}(p)\rangle
\delta({\rm Re}\,^{\ast}\!\Delta^{\!-1\,l}(p)).
\label{eq:4i}
\end{equation}

In order to define the scattering probability ${\it w}_{2n+2}$,
the correlation function on the right-hand side of Eq.\,(\ref{eq:4i})
has to contain terms of $2\,(2n+2)$th order in a free field
$A_{\mu}^{(0)a}$.
The required $2\,(2n + 2)$-order correlator yields the color current
$J_{\mu}^{(s)a}$
(\ref{eq:3o}) for $s=2n + 1$, where the interacting field should be
replaced by free field: $A_{\mu}^a\rightarrow A_{\mu}^{(0)a}$. However, here, it is
also necessary to take into account the effects which arise from iteration of the
currents $J_{\mu}^{(s^{\prime})a}$ of the lower order, $2\!\leq\!s^{\prime}\!<\!2n+1$.
They give a contribution  to the process of $(2n+2)$--plasmon
decay of the same order in coupling constant, as current $J_{\mu}^{(2n+
1)a}$.
The consideration of all
similar contributions can be effectively presented as a replacement of a current
$J^{(s)a}_{\mu}[A]$ by certain effective current
$\tilde{J}^{(s)a}_{\mu}[A^{(0)}]$:
\begin{equation}
J^{(s)a}_{\mu}(A,\ldots,A) \rightarrow
\tilde{J}^{(s)a}_{\mu}(A^{(0)},\ldots,A^{(0)})
\label{eq:4o}
\end{equation}
\[
\equiv \frac{1}{s!}\;g^{s-1}\!\!\int\!\!
\,^{\ast}\tilde{\Gamma}^{a a_1\ldots a_s}_{\mu\mu_1\ldots\mu_s}
(p,-p_{1},\ldots,-p_{s})
\{A^{(0)a_1\mu_1}(p_{1})A^{(0)a_2\mu_2}(p_{2})\ldots A^{(0)a_s\mu_s}(p_{s})
\]
\[
-A^{(0)a_1\mu_1}(p_{1})\langle A^{(0)a_2\mu_2}(p_{2})\ldots
A^{(0)a_s\mu_s}(p_{s})\rangle -\,\ldots\,-
\langle A^{(0)a_1\mu_1}(p_{1})A^{(0)a_2\mu_2}(p_{2})\ldots
A^{(0)a_s\mu_s}(p_{s})\rangle\}
\]
\[
\times\delta (p - \sum_{i=1}^{s}p_{i})\prod_{i=1}^{s}dp_{i},
\]
where the coefficient functions
$\,^{\ast}\tilde{\Gamma}^{a a_1\ldots a_s}_{\mu\mu_1\ldots\mu_s}$
(which in the subsequent discussion we shall call
effective amplitudes as distinct from usual HTL--amplitudes
$\,^{\ast}{\Gamma}^{a a_1\ldots a_s}_{\mu\mu_1\ldots\mu_s}$)
represent highly nontrivial combinations of HTL-amplitudes and an equilibrium
propagator for a soft gluon. Below an algorithm of their succesive definition
will be presented. Here, the fact that a general structure of effective
current $\tilde{J}_{\mu}^{(s)a}$ is similar to a structure of initial
expression (\ref{eq:3o}), is only important for us.

Substituting (\ref{eq:3o}) into Eq.\,(\ref{eq:4i}) we obtain a sum of
products of the correlators of different orders: $2s$th-order correlator,
product of pair correlators multiplied by $(2s - 2)$th-order correlators etc.
The next step is the correlation decoupling of higher correlators in terms of
pairs and next expressing $\langle A^{(0)}A^{(0)}\rangle$ in terms of
plasmon number density $N^l$. However as was shown in Ref.\,\cite{mar3}
the terms in the decomposition do not all
associated with the processes of plasmon decays. The decomposition of averaging
free field amplitudes into the correlators containing the pair of
complex-conjugate potentials or one of the non-conjugate potentials between the
inside of the angular brackets of statistical averaging, does not give
a contribution to the processes of interest to us. The reason of this fact is
similar to reason, by which we have dropped all crossed terms in a product of two
series in initial expression (\ref{eq:4r}). Here, it is not appeared the
requred $\delta$-functions in the phase-space measure, expressing the energy
and the momentum conservation of the decay processes of plasmons. For this
reason in substituting the expression (\ref{eq:4o}) into (\ref{eq:4i})
we can drop all the terms in braces in (\ref{eq:4o}) containing the averaging
and lead to the following expression, instead of (\ref{eq:4i})
\[
{\cal I}^{l\,(s)} = -\pi\!\lim\limits_{\tau,\,V\rightarrow\infty}
\frac{(2\pi)^4}{\tau V}\frac{1}{(s!)^2}\,g^{2s-2}\!
\int\!dp\,\omega\,{\rm sign}(\omega)\delta^{a a^{\prime}}
\tilde{Q}^{\mu\mu^{\prime}}(p)
\,^{\ast}\tilde{\Gamma}^{\dagger\, a a_1\ldots a_s}_{\mu\mu_1\ldots\mu_s}
(p,-p_{1},\ldots,-p_{s})
\]
\[
\times\,^{\ast}\tilde{\Gamma}^{a^{\prime} a^{\prime}_1\ldots
a^{\prime}_s}_{\mu^{\prime}\mu^{\prime}_1\ldots\mu^{\prime}_s}
(p,-p_{1}^{\prime},\ldots,-p_{s}^{\prime})
\langle A^{\ast(0)a_1\mu_1}(p_{1})\ldots A^{\ast(0) a_s\mu_s}(p_{s})
A^{(0)a^{\prime}_1\mu^{\prime}_1}(p^{\prime}_{1})\ldots
A^{(0)a^{\prime}_s\mu^{\prime}_s}(p^{\prime}_{s})\rangle
\]
\begin{equation}
\times\delta^{(4)}(p - \sum_{i=1}^{s}p_{i})
\delta^{(4)}(p - \sum_{i=1}^{s}p^{\prime}_{i})
\,\delta({\rm Re}\,^{\ast}\!\Delta^{\!-1\,l}(p))\!
\prod_{i=1}^{s}dp_{i}dp_{i}^{\prime}.
\label{eq:4p}
\end{equation}
In order not to overburden equations by symbol $`\ast`$ we use sometimes a
dagger $`\dagger`$ to denote complex conjugation.
We write out decoupling of the $2s$th-order correlator on the right-hand side
of Eq.\,(\ref{eq:4p}) in terms of pair correlators, which gives a contribution
to the processes of interest to us. Suppressing color and Lorentz indices and
employing a condensed notion, $A_1\equiv A^{(0)\,a_1}_{{\mu}_1}(p_1)$, we have
\[
\langle A^{\ast(0)}_{1}\ldots A^{\ast(0)}_{s}
A^{(0)}_{1^{\prime}}\ldots A^{(0)}_{s^{\prime}}\rangle =
\Bigl\{\Bigl[
\langle A^{\ast(0)}_{1}A^{(0)}_{1^{\prime}}\rangle\ldots
\langle A^{\ast(0)}_{s}A^{(0)}_{s^{\prime}}\rangle +
({\rm perms.\,of}\,1^{\prime},2^{\prime},\ldots s^{\prime})\Bigr]
\]
\[
+ ({\rm perms.\,of}\,1,2,\ldots s)\Bigr\} =
s!\Bigl\{
\langle A^{\ast(0)}_{1}A^{(0)}_{1^{\prime}}\rangle\ldots
\langle A^{\ast(0)}_{s}A^{(0)}_{s^{\prime}}\rangle +
({\rm perms.\,of}\,1^{\prime},2^{\prime},\ldots s^{\prime})\Bigr\}.
\]
All terms on the right-hand side of the latter equality, that is obtaned from
the first one by all possible permutations of arguments
$(p_i^{\prime},a_i^{\prime},\mu_i^{\prime}),\; i=1,\ldots,s$,
give the same contribution to the emitted
radiant power ${\cal I}^{l\,(s)}$ and therefore, here, we can write
\[
\langle A^{\ast(0)}_{1}\ldots A^{\ast(0)}_{s}
A^{(0)}_{1^{\prime}}\ldots A^{(0)}_{s^{\prime}}\rangle =
(s!)^2
\langle A^{\ast(0)}_{1}A^{(0)}_{1^{\prime}}\rangle\ldots
\langle A^{\ast(0)}_{s}A^{(0)}_{s^{\prime}}\rangle
\]
\[
=(s!)^2\prod_{i=1}^{s}\delta^{a_i a_i^{\prime}}
\tilde{Q}^{\mu_i \mu_i^{\prime}}(p_i) I^l(p_i)\delta(p_i - p_i^{\prime}).
\]
Here, in the last line we use a definition of the correlation function
$I^{\mu_i \mu_i^{\prime}}(p_i)$ in the equilibrium, Eq.\,(\ref{eq:3f}), and
leave only the longitudinal part in its expansion (\ref{eq:3h}). Now we
substitute the obtained correlation decoupling into Eq.\,(\ref{eq:4p}) and
perform an integration over $\prod_{i=1}^{s}dp_i^{\prime}$,
\begin{equation}
{\cal I}^{l(s)} = -\pi\!\lim\limits_{\tau,\,V\rightarrow\infty}
\frac{1}{\tau V}\,g^{2s-2}\!
\int\!dp\!\int\prod_{i=1}^{s}dp_i\,(2\pi)^4
\Bigr[\delta^{(4)}(p - \sum_{i=1}^{s}p_{i})\Bigl]^2
\omega\,{\rm sign}(\omega)\,\tilde{Q}^{\mu\mu^{\prime}}(p)
\label{eq:4a}
\end{equation}
\[
\times\prod_{i=1}^{s}\tilde{Q}^{\mu_i\mu_i^{\prime}}(p_i)
\!\,^{\ast}\tilde{\Gamma}^{\dagger\, a a_1\ldots a_s}_{\mu\mu_1\ldots\mu_s}
(p,-p_{1},\ldots,-p_{s})
\!\,^{\ast}\tilde{\Gamma}^{a a_1\ldots a_s}_{\mu^{\prime}\mu^{\prime}_1
\ldots\mu^{\prime}_s}(p,-p_{1},\ldots,-p_{s})
\prod_{i}^{s}I^l(p_i)
\delta({\rm Re}\,^{\ast}\!\Delta^{\!-1\,l}(p)).
\]
By a $\delta$-functions squared here, we mean as usual \cite{bjo}
\[
\Bigr[\delta^{(4)}(p - \sum_{i=1}^{s}p_{i})\Bigl]^2 =
\frac{1}{(2\pi)^4}\,\tau V\delta(p-\sum_{i=1}^{s}p_{i}).
\]

To take into account weakly inhomogeneous and weakly non-stationary in the
medium it is sufficient, within the accepted accuracy, to replace
equilibrium spectral densities $I^l(p_i)$ by off-equlibrium ones in the
Wigner form (\ref{eq:3g}): $I^l(p_i)\rightarrow I^l(p_i,x_i)$, slowly depending
on $x$. Futhermore, we take the functions $I^l(p_i,x_i)$ in the form of
{\it the quasiparticle approximation}
\begin{equation}
I^l(p_i,x_i) = I_{{\bf p}_i}^l \delta (\omega_i - \omega_{{\bf p}_i}^l)
+ I_{-{\bf p}_i}^l \delta (\omega_i + \omega_{{\bf p}_i}^l), \quad
I_{{\bf p}_i}^l\equiv I^l({\bf p}_i,x_i)
\label{eq:4s}
\end{equation}
and goes over from functions $I_{{\bf p}_i}^l$ to the plasmon number density
\begin{equation}
N_{{\bf p}_i}^l = -\,(2\pi)^3\,2\omega_{{\bf p}_i}^l {\rm Z}_l^{-1}({\bf p}_i)
I_{{\bf p}_i}^l.
\label{eq:4d}
\end{equation}
Since external soft-gluon legs lie on the plasmon mass-shell, then account
must be taken of the fact that conservation laws of energy and momentum
(see, section 2) admit decay processes with even number of plasmons, i.e.
it is necessary to set $s=2n+1,\;n=1,2,\ldots$.

Taking into account above-mentioned, and Eqs.\,(\ref{eq:3s}), (\ref{eq:4y}),
(\ref{eq:4s}) and (\ref{eq:4d}) we obtain instead of (\ref{eq:4a})
\[
{\cal I}^{l\,(2n+1)} = \frac{1}{2}
\int\!d\omega\!\int \prod_{i=1}^{2n+1}\!d\omega_i\!
\int\!\frac{d{\bf p}}{(2\pi)^3}\!\int\prod_{i=1}^{2n+1}
\frac{d{\bf p}_i}{(2\pi)^3}\,
(2\pi)^4\delta^{(4)}(p - \sum_{i=1}^{2n+1}p_{i})\,
\omega\,{\rm sign}(\omega)
\]
\begin{equation}
\times{\rm M}^{\ast\,a a_1 \ldots a_{2n+1}}
({\bf p}, -{\bf p}_1,\ldots,-{\bf p}_{2n+1})
{\rm M}^{a a_1 \ldots a_{2n+1}}
({\bf p}, -{\bf p}_1,\ldots,-{\bf p}_{2n+1})
\label{eq:4f}
\end{equation}
\[
\times\{\delta (\omega - \omega_{{\bf p}}^l)
+ \delta (\omega + \omega_{{\bf p}}^l)\}
\prod_{i=1}^{2n+1}\{N_{{\bf p}_i}^l \delta (\omega_i - \omega_{{\bf p}_i}^l)
+ N_{-{\bf p}_i}^l \delta (\omega_i + \omega_{{\bf p}_i}^l)\}.
\]
Here, we introduce
\begin{equation}
{\rm M}^{a a_1 \ldots a_{2n+1}}
({\bf p}, -{\bf p}_1,\ldots,-{\bf p}_{2n+1})= g^{2n}\!
\left(\frac{{\rm Z}_l({\bf p})}{2\omega_{\bf p}^l}\right)^{1/2}
\Biggl(\frac{\tilde{u}^{\mu}(p)}{\sqrt{\bar{u}^2(p)}}\Biggr)
\label{eq:4g}
\end{equation}
\[
\times\prod_{i=1}^{2n+1}
\left(\frac{{\rm Z}_l({\bf p}_i)}{2\omega_{{\bf p}_i}^l}\right)^{1/2}
\biggl(\frac{\tilde{u}^{\mu_i}(p_i)}{\sqrt{\bar{u}^2(p_i)}}\biggr)
\,^{\ast}\tilde{\Gamma}^{a a_1\ldots a_{2n+1}}_{\mu\mu_1\ldots\mu_{2n+1}}
(p,-p_{1},\ldots,-p_{2n+1})\mid_{\rm on-shell}
\]
representing the interaction matrix element for $(2n+2)$--plasmon decay. 4-vectors
\[
\left(\frac{{\rm Z}_l({\bf p})}{2\omega_{\bf p}^l}\right)^{1/2}
\!\frac{\tilde{u}^{\mu}(p)}{\sqrt{\bar{u}^2(p)}}\equiv\frac{1}
{\sqrt{2\omega^l_{\bf p}}}\,\epsilon^{\mu}(p,\lambda),\quad\lambda=l
\]
on the right-hand side of (\ref{eq:4g}) represent usual wave functions of
longitudinal physical gluon in temporal gauge, where factor
${\rm Z}_l^{1/2} ({\bf p})$ provides renormalization of gluon wave
function by thermal effects.

Furthermore we multiply out terms in curly brackets in integrand of
Eq.\,(\ref{eq:4f}). In terms containing the factor of
$N_{- {\bf p}_i}^l \delta (\omega_i+\omega_{{\bf p}_i}^l)$ type we replace
${\bf p}_i \rightarrow - {\bf p}_i$ (in so doing we follow the rule
$\omega_{- {\bf p}_i}^l \rightarrow -\,\omega_{{\bf p}_i}^l$). By arguments
presented in section 2 it should be left only the terms responsible for
elastic plasmon scattering of (\ref{eq:2u}) type, and drop the remaining
ones in obtained expression. In this case integrand in Eq.\,(\ref{eq:4f}) has
the form
\[
2(2\pi)^4\omega_{\bf p}^l\Bigl\{\delta^{(4)}
(p+p_1+\ldots+p_n-p_{n+1} -\ldots -p_{2n+1})
{\rm M}^{\ast \{a\}}({\bf p},{\bf p}_1,\ldots,{\bf p}_n,-{\bf p}_{n+1},
\ldots, -{\bf p}_{2n+1})
\]
\begin{equation}
\times{\rm M}^{\{a\}}
({\bf p},{\bf p}_1,\ldots,{\bf p}_n,-{\bf p}_{n+1},\ldots, -{\bf p}_{2n+1})
+ ({\rm perms.\,of\,signs})\Bigr\}\!\prod_{i=1}^{2n+1}\!N_{{\bf p}_i}^l.
\label{eq:4h}
\end{equation}
Here, for brevity we enter multi-index notation
$\{a\} = (a, a_1, \ldots, a_{2n + 1})$ and fix a sign of a first argument of the
matrix element ${\rm M}^{\{a\}}$. The factor 2 takes into account a contribution
of the term $\delta(\omega + \omega_{\bf p}^l)$. The permutation over sign
means all possible placements of $n + 1$ minus signs among
external momenta ${\bf p}_1, {\bf p}_2,\ldots{\bf p}_{2n + 1}$. As the last
step it should be performed a replacement of momenta such that arguments of
$\delta$-functions for all terms in braces of Eq.\,(\ref{eq:4h}) become equal to
argument of $\delta$-function entering into the first term. This enables us
put outside the brackets of $\delta$-function as a common factor.
Confronting an obtained expression for ${\cal I}^{l\,(2n + 1)}$ and
corresponding term in expansion (\ref{eq:4q}), one identifies the required
probability ${\it w}_{2n + 2}$:
\begin{equation}
{\it w}_{2n+2}
({\bf p},{\bf p}_1,\ldots,{\bf p}_n;{\bf p}_{n+1},\ldots,{\bf p}_{2n+1})
\label{eq:4j}
\end{equation}
\[
= {\rm M}^{\ast \{a\}}({\bf p},{\bf p}_1,\ldots,{\bf p}_n,-{\bf p}_{n+1},
\ldots, -{\bf p}_{2n+1})
{\rm M}^{\{a\}}
({\bf p},{\bf p}_1,\ldots,{\bf p}_n,-{\bf p}_{n+1},\ldots, -{\bf p}_{2n+1})
\]
\[
+\sum_{1\leq i_1\leq n}^{({\bf p_i})}{\rm M}^{\ast\{a\}}{\rm M}^{\{a\}}
+\sum_{1\leq i_1<i_2\leq n}^{({\bf p}_{i_1},\,{\bf p}_{i_2})}
{\rm M}^{\ast\{a\}}{\rm M}^{\{a\}}+
\ldots+\sum^{({\bf p}_1,\ldots,{\bf p}_n)}
{\rm M}^{\ast\{a\}}{\rm M}^{\{a\}},
\]
where on the last line ${\rm M}^{\{a\}}\equiv
{\rm M}^{\{a\}}
({\bf p},{\bf p}_1,\ldots,{\bf p}_n,-{\bf p}_{n+1},\ldots, -{\bf p}_{2n+1})$.
The summing symbol $\sum\limits_{1\leq i_1\leq n}^{({\bf p_i})}$
denotes summing over all possible momentum interchange
${\bf p}_{i_1},\;1\leq i_1\leq n$ by momentum $(-{\bf p}_{j_1}),\;
n+1\leq j_1\leq 2n+1$. The symbol
$\sum_{1\leq i_1<i_2\leq n}^{({\bf p}_{i_1},{\bf p}_{i_2})}$ analogously
denotes a summing over all possible interchange of momenta pair
$({\bf p}_{i_1},{\bf p}_{i_2}),\;1\leq i_1\!<\!i_2\leq n$ by momenta pair
$(-{\bf p}_{j_1},-{\bf p}_{j_2})$, where $n+1\leq j_1\!<\!j_2\leq 2n+1$ etc.
Let us consider as an example a case of four-plasmon decay, that corresponds to
$n=1$. By Eq.\,(\ref{eq:4j}) the scattering probability ${\it w}_4$ has a form
\begin{equation}
{\it w}_4({\bf p},{\bf p}_1;{\bf p}_2,{\bf p}_3)=
{\rm M}^{\ast\{a\}}({\bf p},{\bf p}_1,-{\bf p}_2,-{\bf p}_3)
{\rm M}^{\{a\}}({\bf p},{\bf p}_1,-{\bf p}_2,-{\bf p}_3)
\label{eq:4k}
\end{equation}
\[
+\,{\rm M}^{\ast\{a\}}({\bf p},-{\bf p}_2,{\bf p}_1,-{\bf p}_3)
{\rm M}^{\{a\}}({\bf p},-{\bf p}_2,{\bf p}_1,-{\bf p}_3)
\]
\[
+\,{\rm M}^{\ast\{a\}}({\bf p},-{\bf p}_3,-{\bf p}_2,{\bf p}_1)
{\rm M}^{\{a\}}({\bf p},-{\bf p}_3,-{\bf p}_2,{\bf p}_1).
\]
Here, three terms imply the existence of three possible channels of this
process, which change the plasmon number density $N_{\bf p}^l$
\[
{\rm g}^{\ast}+{\rm g}_1^{\ast}
\rightleftharpoons{\rm g}_2^{\ast}+{\rm g}_3^{\ast},\quad
{\rm g}^{\ast}+{\rm g}_2^{\ast}
\rightleftharpoons{\rm g}_1^{\ast}+{\rm g}_3^{\ast},\quad
{\rm g}^{\ast}+{\rm g}_3^{\ast}
\rightleftharpoons{\rm g}_1^{\ast}+{\rm g}_2^{\ast}.
\]
The expression (\ref{eq:4k}) is suitable for checking on the symmetry conditions
(\ref{eq:2i}), which is imposed on the plasmon-plasmon scattering probability.\\

\section{\bf The Boltzmann equation for four-colorless plasmon decays}
\setcounter{equation}{0}

In this section, we review the main features of the Boltzmann equation for
four-colorless plasmon decays which recently derived in Ref.\,\cite{mar3}.
There are no
new results to be reported here, and they are given for coherent and complete
subsequent discussion. By general formula (\ref{eq:4g}) the matrix element
${\rm M}^{a a_1 a_2 a_3} ({\bf p}, {\bf p}_1, {\bf p}_2, {\bf p}_3)$ of four-plasmon
decay is defined by the effective amplitude
$^{\ast} \tilde{\Gamma}_{\mu \mu_1 \mu_2 \mu_3}^{a a_1 a_2 a_3}
(p, p_1, p_2, p_3)$ that in turn is a coefficient function in integrand of
an effective current $\tilde{J}_{\mu}^{(3)a}[A^{(0)}]$ (Eq.\,(\ref{eq:4o})).
Such a problem is reduced to a definition of explicit form of
$\tilde{J}_{\mu}^{(3)a}[A^{(0)}]$. For this purpose we use a
basic field equation (\ref{eq:3i}), where on the right-hand side we
leave the terms up to the third order in powers of $A_{\mu}$
\begin{equation}
\,^{\ast}\tilde{\cal D}^{-1 \, \mu \nu}(p) A^{a}_{\nu}(p) =
-J^{(2)a \mu}(A,A) - J^{(3)a \mu}(A,A,A),
\label{eq:5q}
\end{equation}
where,
\[
J^{(2)a}_{\mu}(A,A) =\frac{1}{2!}\,g\!\!\int\!
\,^{\ast}\Gamma^{a a_1a_2}_{\mu \mu_1 \mu_2}(p,-p_{1},-p_{2})
\{A^{a_1\mu_1}(p_{1})A^{a_2\mu_2}(p_{2})-
\langle A^{a_1\mu_1}(p_{1})A^{a_2\mu_2}(p_{2})\rangle)\}
\]
\begin{equation}
\times\delta (p - p_{1} - p_{2}) dp_{1}dp_{2},
\label{eq:5w}
\end{equation}
and
\[
J^{(3)a}_{\mu}(A,A,A) =\frac{1}{3!}\,g^2\!\!\int\!
\,^{\ast}\Gamma^{a a_1a_2a_3}_{\mu \mu_1 \mu_2\mu_3}(p,-p_{1},-p_{2},-p_3)
\{A^{a_1\mu_1}(p_{1})A^{a_2\mu_2}(p_{2})A^{a_3\mu_3}(p_{3})
\]
\[
-A^{a_1\mu_1}(p_{1})\langle A^{a_2\mu_2}(p_{2})A^{a_3\mu_3}(p_{3})\rangle
-\langle A^{a_1\mu_1}(p_{1})A^{a_2\mu_2}(p_{2})A^{a_3\mu_3}(p_{3})\rangle
\}\delta (p - p_{1} - p_{2}-p_{3}) dp_{1}dp_{2}dp_{3}.
\]
The function
\[
\,^{\ast}\Gamma^{a a_1a_2}_{\mu \mu_1 \mu_2}(p,-p_{1},-p_{2})=
-if^{aa_1a_2}\,^{\ast}\Gamma_{\mu \mu_1 \mu_2}(p,-p_{1},-p_{2}),
\]
\[
\,^{\ast}\Gamma^{a a_1a_2a_3}_{\mu \mu_1 \mu_2\mu_3}(p,-p_{1},-p_{2},-p_3)=
-f^{aa_1b}f^{ba_2a_3}
\,^{\ast}\Gamma_{\mu\mu_1\mu_2\mu_3}(p,-p_{1},-p_{2},-p_3)
\]
\[
-\,f^{aa_2b}f^{ba_1a_3}
\,^{\ast}\Gamma_{\mu\mu_2\mu_1\mu_3}(p,-p_{2},-p_{1},-p_3),
\]
are three and four-gluon HTL-amplitudes accordingly.

The nonlinear integral equation (\ref{eq:5q}) is solved by the approximation
scheme method. Discarding the nonlinear terms in $A$ on the right-hand side
of Eq.\,(\ref{eq:5q}), we obtain in the first approximation
\[
\,^{\ast}\tilde{\cal D}^{-1 \, \mu \nu}(p) A^{a}_{\nu}(p) = 0.
\]
The solution of this equation, which we denote by $A^{(0)a}_{\mu}(p)$, is
the solution for a free field.

Furthermore keeping the term, quadratic in field on the right-hand side of
Eq.\,(\ref{eq:5q}), we derive
\[
\,^{\ast}\tilde{\cal D}^{-1 \, \mu \nu}(p) A^{a}_{\nu}(p) =
-J^{(2)a \mu}(A^{(0)},A^{(0)}),
\]
where on the right-hand side we substitute free fields instead of interacting
ones. The general solution of the last equation is given in the form
\[
A^{a}_{\mu}(p) = A^{(0)a}_{\mu}(p)
- \,^{\ast}\tilde{\cal D}_{\mu \nu}(p)\tilde{J}^{(2)a \nu}(A^{(0)},A^{(0)}),
\]
where $\tilde{J}^{(2)a \nu}(A^{(0)},A^{(0)})\equiv
J^{(2)a \nu}(A^{(0)},A^{(0)})$.

The following term in the expansion of the interacting field is defined from
equation
\begin{equation}
\,^{\ast}\tilde{\cal D}^{-1 \, \mu \nu}(p) A^{a}_{\nu}(p) =
-J^{(2)a \mu}(- \,^{\ast}\tilde{\cal D}J^{(2)}(A^{(0)},A^{(0)}),A^{(0)})
\label{eq:5e}
\end{equation}
\[
-\,J^{(2)a \mu}(A^{(0)},- \,^{\ast}\tilde{\cal D}J^{(2)}(A^{(0)},A^{(0)}))
-\, J^{(3)a \mu}(A^{(0)},A^{(0)},A^{(0)}).
\]
The first two terms on the right-hand side of this equation represent
iteration of the currents $J_{\mu}^{(s^{\prime})\,a}$
of lower order (in this case, $s^{\prime}=2$) mentioned in previous section,
whose contribution is of the same order in the coupling $g$
as $J^{(3)a\mu}(A^{(0)},A^{(0)},A^{(0)})$.
Using explicit expressions for currents (\ref{eq:5w}),
after cumbersome algebraic transformations, we obtain the form of interacting
field from the Eq. (\ref{eq:5e}) with accuracy required for our further calculations
\[
A^{a}_{\mu}(p) = A^{(0)a}_{\mu}(p)
- \,^{\ast}\tilde{\cal D}_{\mu \nu}(p)\tilde{J}^{(2)a \nu}(A^{(0)},A^{(0)})-
\,^{\ast}\tilde{\cal D}_{\mu \nu}(p)\tilde{J}^{(3)a\nu}(A^{(0)},A^{(0)},A^{(0)}).
\]
Here, third-order color current on the right-hand side is defined by expression
\[
\tilde{J}^{(3)a}_{\mu}(A^{(0)},A^{(0)},A^{(0)})
\]
\begin{equation}
= \frac{1}{3!}\,g^{2}\!\!\int\!
\,^{\ast}\tilde{\Gamma}^{a a_1 a_2 a_3}_{\mu\mu_1\mu_2\mu_3}
(p,-p_{1},-p_{2},-p_{3})
\{A^{(0)a_1\mu_1}(p_{1})A^{(0)a_2\mu_2}(p_{2})A^{(0)a_3\mu_3}(p_{3})
\label{eq:5r}
\end{equation}
\[
-A^{(0)a_1\mu_1}(p_{1})\langle A^{(0)a_2\mu_2}(p_{2})
A^{(0)a_3\mu_3}(p_{3})\rangle
- \langle A^{(0)a_1\mu_1}(p_{1})A^{(0)a_2\mu_2}(p_{2})
A^{(0)a_3\mu_3}(p_{3})\rangle\}
\]
\[
\times\delta (p - p_1 - p_2 - p_3)dp_1dp_2dp_3,
\]
where the effective four-gluon amplitude
$\,^{\ast}\tilde{\Gamma}^{a a_1 a_2 a_3}_{\mu\mu_1\mu_2\mu_3}
(\equiv\,^{\ast}\tilde{\Gamma}^{(4)})$
is defined by expression
\begin{equation}
\,^{\ast}\tilde{\Gamma}^{aa_1a_2a_3}_{\mu\mu_1\mu_2\mu_3}(p,-p_{1},-p_{2},-p_3)=
-f^{aa_1b}f^{ba_2a_3}
\,^{\ast}\tilde{\Gamma}_{\mu \mu_1 \mu_2\mu_3}(p,-p_{1},-p_{2},-p_3)
\label{eq:5t}
\end{equation}
\[
-\,f^{aa_2b}f^{ba_1a_3}
\,^{\ast}\tilde{\Gamma}_{\mu \mu_2 \mu_1\mu_3}(p,-p_{2},-p_{1},-p_3).
\]
The color factors in Eq.\,(\ref{eq:5t}) are multiplied by purely kinematical
coefficients, which we shall call partial effective amplitudes or effective
subamplitudes\footnote{Here, we use the terminology accepted in theory
of multi-parton hard processes (see, e.g. \cite{ber}), where the expansion of
(\ref{eq:5t}) type in calculation of the tree QCD amplitudes in the high-energy
limit, are also arizen.}, and defined as follows
\[
\,^{\ast}\tilde{\Gamma}_{\mu \mu_1 \mu_2\mu_3}(p,-p_{1},-p_{2},-p_3)
\equiv
\,^{\ast}{\Gamma}_{\mu \mu_1 \mu_2\mu_3}(p,-p_{1},-p_{2},-p_3)
\]
\begin{equation}
-\!\,^{\ast}\Gamma_{\mu\mu_1\nu}(p,-p_1,-p+p_1)
\,^{\ast}\!\tilde{\cal D}^{\nu\nu^{\prime}}(p_2+p_3)
\,^{\ast}\Gamma_{\nu^{\prime}\mu_2\mu_3}(p_2+p_3,-p_2,-p_3)
\label{eq:5y}
\end{equation}
\[
- \,^{\ast}\Gamma_{\mu\mu_3\nu}(p,-p_3,-p+p_3)
\,^{\ast}\!\tilde{\cal D}^{\nu\nu^{\prime}}(p_1+p_2)
\,^{\ast}\Gamma_{\nu^{\prime}\mu_2\mu_1}(p_1+p_2,-p_2,-p_1).
\]
For complete picture we leave the term
$\langle A^{(0)} A^{(0)} A^{(0)}\rangle$
in braces under the integral sign on the right-hand side of Eq.\,(\ref{eq:5r}).
It is clear that this term
is equal to zero, because $A^{(0)}$ represents the amplitudes of entirely
uncorrelated waves. The expression (\ref{eq:5r}) represents desired effective
current $\tilde{J}_{\mu}^{(3)a} [A^{(0)}]$, whose coefficient function by a
formula (\ref{eq:4g}) defines the matrix element for four-plasmon decay
\begin{equation}
{\rm M}^{a a_1 a_2 a_3}
({\bf p}, -{\bf p}_1,-{\bf p}_2,-{\bf p}_3)= g^{2}\!
\left(\frac{{\rm Z}_l({\bf p})}{2\omega_{\bf p}^l}\right)^{1/2}
\!\Biggl(\frac{\tilde{u}^{\mu}(p)}{\sqrt{\bar{u}^2(p)}}\Biggr)
\label{eq:5u}
\end{equation}
\[
\times\prod_{i=1}^{3}
\left(\frac{{\rm Z}_l({\bf p}_i)}{2\omega_{{\bf p}_i}^l}\right)^{1/2}
\!\Biggl(\frac{\tilde{u}^{\mu_i}(p_i)}{\sqrt{\bar{u}^2(p_i)}}\Biggr)
\,^{\ast}\tilde{\Gamma}^{a a_1 a_2 a_3}_{\mu\mu_1\mu_2\mu_3}
(p,-p_{1},-p_2, -p_3)\Big|_{\rm on-shell}.
\]
The obtained matrix element has a simple diagrammatic representation
drawn in Fig.\,1. The black square here, denotes an effective amplitude
$^{\ast}\tilde{\Gamma}^{(4)}$.
The first term on the right-hand side Fig.\,1 represents a direct interaction of
four plasmons, that induced by usual four-gluon HTL-amplitude. The remaining
terms are connected with plasmons interaction, induced by three-gluon
HTL-amplitudes with intermediate virtual oscillations representing $s$- and
$t$-channel contributions respectively (effective subamplitude
$^{\ast}\tilde{\Gamma}_{\mu \mu_2 \mu_1\mu_3}(p,-p_{2},-p_{1},-p_3)$ in
Eq.\,(\ref{eq:5t}) contains also $u$-channel contribution, which we
have not drawn in Fig.\,1).

With the matrix element (\ref{eq:5u}) in hand it is not difficult to write
the Boltzmann equation describing the plasmon-plasmon scattering. Here,
we have restricted our consideration to linearized version of this equation. For
this purpose we assume that the off-equilibrium fluctuation is perturbative small
and write the number density of colorless plasmons as
\[
N^l_{\bf p} = N_{eq}^l({\bf p}) + \delta N^l_{\bf p},
\]
where $N_{eq}^{l}({\bf p}) = (e^{\omega^{l}_{\bf p}/T^{\ast}} - 1)^{-1}$
is the Planck distribution function and $T^{\ast}$ is a certain constant, which can be interpreted as a
plasmon gas temperature in the statistical equilibrium state.
Parametrizating  off-equilibrium fluctuation of the occupation number
$\delta N^l_{\bf p}$ as follows
\begin{equation}
\delta N^l_{\bf p} \equiv - \frac{dN^l_{eq}({\bf p})}
{d\omega^l_{\bf p}}\,{\cal W}^l_{\bf p} =
\frac{1}{T^{\ast}}\,N^l_{eq}({\bf p})(N^l_{eq}({\bf p})+1){\cal W}^l_{\bf p},
\label{eq:5i}
\end{equation}
we derive from Eqs.\,(\ref{eq:2w})\,--\,(\ref{eq:2t}) for $n=1$, and
Eqs.\,(\ref{eq:4k}) and (\ref{eq:5u}), after simple color algebra,
a linearized Boltzmann equation for function ${\cal W}^l_{\bf p}$
\[
\frac{\partial {\cal W}^l_{\bf p}}{\partial t} +
{\bf V}^l_{\bf p}\cdot
\frac{\partial {\cal W}^l_{\bf p}}{\partial {\bf x}} =
-\!\int\!
\frac{d{\bf p}_1}{(2\pi)^3}\frac{d{\bf p}_2}{(2 \pi)^3}
\frac{d{\bf p}_3}{(2 \pi)^3} \,
(2 \pi)^4\delta (\omega_{\bf p}^l+\omega_{{\bf p}_1}^l-\omega_{{\bf p}_2}^l -
\omega_{{\bf p}_3}^l)
\]
\begin{equation}
\times\!\delta({\bf p}+{\bf p}_1 - {\bf p}_2 - {\bf p}_3)\,
\frac{N_{eq}^l({\bf p}_1)(N_{eq}^l({\bf p}_2) + 1)
(N_{eq}^l({\bf p}_3) + 1)}
{(N_{eq}^l({\bf p}) + 1)}
\label{eq:5o}
\end{equation}
\[
\times{\it w}({\bf p},{\bf p}_1;{\bf p}_2,{\bf p}_3)
\{{\cal W}^l_{\bf p} - {\cal W}^l_{{\bf p}_2} +
{\cal W}^l_{{\bf p}_1} - {\cal W}^l_{{\bf p}_3}\},
\]
where the function
\[
{\it w}({\bf p},{\bf p}_1;{\bf p}_2,{\bf p}_3)= 3g^4N_c^2
\left(\frac{{\rm Z}_l({\bf p})}{2\omega_{\bf p}^l\bar{u}^2(p)}\right)
\prod_{i=1}^{3}
\left(\frac{{\rm Z}_l({\bf p}_i)}{2\omega_{{\bf p}_i}^l\bar{u}^2(p_i)}\right)
\{\vert \,^{\ast} \tilde{\Gamma} (p, - p_2, p_1, - p_3) \vert^2
\]
\begin{equation}
+\,\vert \,^{\ast} \tilde{\Gamma} (p, p_1, - p_3, - p_2) \vert^2
+ {\rm Re} \,( \,^{\ast} \tilde{\Gamma}(p, - p_2, p_1, - p_3)
\,^{\ast} \tilde{\Gamma}^{\dagger}(p, p_1, - p_3, - p_2)) \}\vert_{\rm on-shell}
\label{eq:5p}
\end{equation}
is the probability of four-colorless plasmon decay. Here, we denote
\begin{equation}
\,^{\ast}\tilde{\Gamma} (p, p_1, p_2, p_3) \equiv
\,^{\ast}\tilde{\Gamma}^{\mu\mu_1\mu_2\mu_3} (p, p_1, p_2, p_3)
\tilde{u}_{\mu}(p)\tilde{u}_{{\mu}_1}(p_1)\tilde{u}_{{\mu}_2}(p_2)
\tilde{u}_{{\mu}_3}(p_3).
\label{eq:5a}
\end{equation}
The structure of the expression on the right-hand side of Eq.\,(\ref{eq:5o})
is just as the structure of the expression for the case of the
Boltzmann equation for hard gluons \cite{bla3}. However, here,
we have more complicated
and cumbersome expression for scattering probability of two soft quasiparticles
involving the resummed vertices.
Besides, the momentum transfer ${\bf q}={\bf p}-{\bf p}_2={\bf p}_3 -
{\bf p}_1$ is of the same order as the momentum of the soft quasiparticles. In
the last case it is impossible to expand functions
${\cal W}^l_{{\bf p}_2 = {\bf p} - {\bf q}}$ and
${\cal W}^l_{{\bf p}_3 = {\bf p}_1 + {\bf q}}$ as power series in momentum
transfer ${\bf q}$, as it takes place for hard gluons \cite{bla3}.

Using the linearized Boltzmann equation (\ref{eq:5o}) it is not difficult to
estimate the order of the lifetimes of plasmon $\tau^l$ caused by the process of
four-plasmon decay. Considering the plasmon gas in thermal equilibrium with
hard particles from the heat bath, i.e. $T^{\ast}\simeq T$, and using
Eqs.\,(\ref{eq:5o}) and (\ref{eq:5p}), we obtain
\[
\frac{1}{\tau^l} \sim g^3 N_c^2 T.
\]
Thus necessary condition $\omega^l\tau^l\gg 1$ (section 2), such that
excited state of plasma can be described in terms of the plasmon occupation
number $N_{\bf p}^l$ for $g\ll 1$, is high accurate, since here, we have
$\omega^l \tau^l \sim 1/g^{2}$.

At the end of this section we also represent some properties of the effective
subamplitude $\,^{\ast}\tilde{\Gamma}_{\mu\mu_1\mu_2\mu_3} (p, -p_1, -p_2, -p_3)$
defined by Eq.\,(\ref{eq:5y}). In particular, deriving Eqs.\,(\ref{eq:5o})
and (\ref{eq:5p}) we have used two relations which satisfy an effective subamplitude
\begin{equation}
\,^{\ast} \tilde{\Gamma}_{\mu\mu_1\mu_2\mu_3}(p, -p_1, -p_2, -p_3) +
\,^{\ast} \tilde{\Gamma}_{\mu\mu_2\mu_1\mu_3} (p, -p_2,-p_1, -p_3) +
\,^{\ast} \tilde{\Gamma}_{\mu\mu_1\mu_3\mu_2} (p, -p_1, -p_3, -p_2) = 0,
\label{eq:5s}
\end{equation}
\begin{equation}
\,^{\ast}\tilde{\Gamma}_{\mu\mu_1\mu_2\mu_3}(p, -p_1, -p_2, -p_3)=
\,^{\ast}\tilde{\Gamma}_{\mu\mu_3\mu_2\mu_1}(p, -p_3, -p_2, -p_1).
\label{eq:5d}
\end{equation}
Their correctness may be verified by direct calculation using known properties
of HTL-amplitudes \cite{bra1}, entering into the definition of
$\,^{\ast}\tilde{\Gamma}^{(4)}$. These relations show that the effective amplitude
$^{\ast} \tilde{\Gamma}^{(4)}$ possesses some properties of usual four-gluon
HTL-amplitude. As we shall
show below this statement takes place for the effective amplitudes
$^{\ast} \tilde{\Gamma}^{(s)}$ for any values $s \geq 4$
(the case when $s=3$ is trivial by virtue of
$^{\ast}\tilde{\Gamma}^{(3)}\equiv\Gamma^{(3)}$).
Here, the exception, for example, is the Ward identity,
which for $^{\ast} \tilde{\Gamma}^{(4)}$ has not a standard form.\\

\section{\bf Characteristic amplitudes of the soft gluon field}
\setcounter{equation}{0}

In this section we shall estimate for which typical amplitude of the soft
gluon field, the contribution of four-plasmon decay process to
a generalized decay and regeneration rates,
$\Gamma_{\rm d} [N_{\bf p}^l]$ and $\Gamma_{\rm i} [N_{\bf p}^l]$, will be
leading and for which value all terms in the expansions (\ref{eq:2e})
will be of the same order in $g$. At the last case two problem arise:
the computation of explicit form of matrix elements of decay processes of
high orders, for $n>1$, and the problem of summing (\ref{eq:2e}). Recall
that in this work we consider the collective processes only at the soft
momentum scale for $\omega,\vert{\bf p}\vert\sim gT$.

First of all we estimate an order of matrix element
${\rm M}^{a a_1 \ldots a_{2n + 1}}$. We use general expression
(\ref{eq:4g}) connecting matrix element for $(2n + 2)$-plasmon decay with
the effective amplitude $^{\ast}\tilde{\Gamma}^{(2n+2)}$ in which the
kinematical factors corresponding to external soft-gluon legs are single out.
In the soft region of the momentum scale the following estimation results
from this expression
\[
{\rm M}^{a a_1 \ldots a_{2n + 1}} \sim (gT)^{-(n+1)} \,
^{\ast}\tilde{\Gamma}_{\mu\mu_1\ldots\mu_{2n+1}}^{a a_1\ldots a_{2n + 1}}.
\]
Here, the symbol $\sim$ denotes that the order of value from the left in
coupling constant equals to the order of value from the right. The order of
effective amplitude
$^{\ast}\tilde{\Gamma}_{\mu\mu_1\ldots\mu_{2n+1}}^{a a_1\ldots a_{2n + 1}}$
can be estimated from arbitrary tree diagram
with amputate $2n+2$ soft external legs. Let us consider, for example,
the diagram drawn in Fig.\,2. Four-gluon vertices are of order $g^2$, and
HTL-resummed propagators are of order $1/(gT)^2$.
From simple power counting of the diagram it follows an estimation
\[
^{\ast}\tilde{\Gamma}_{\mu\mu_1\ldots\mu_{2n+1}}^{a a_1\ldots a_{2n + 1}}
\sim g^{2n}N_c^n\,\frac{1}{(gT)^{2(n-1)}},
\]
and thus
\begin{equation}
{\rm M}^{a a_1 \ldots a_{2n + 1}}\sim g^{2n}N_c^n\,\frac{1}{(gT)^{3n-1}}.
\label{eq:6q}
\end{equation}
Furthermore integration measure has an estimation
\begin{equation}
d{\cal T}^{(2n+1)}\sim(gT)^{6n-1}.
\label{eq:6w}
\end{equation}
Let us define an order in $g$ of the plasmon density. For this
purpose we use initial relations connecting the plasmon number density
$N_{\bf p}^l$ with two-point correlation function (\ref{eq:3d}) of
soft-gluon field. Here we have
\[
N_{{\bf p}}^l = -\,(2\pi)^3\,2\omega_{{\bf p}}^l {\rm Z}_l^{-1}({\bf p})
I_{{\bf p}}^l,
\]
where in turns the spectral density $I_{{\bf p}}^l$ is defined from relation
$I^l_p = I_{{\bf p}}^l\delta(\omega - \omega_{{\bf p}}^l)
+ I_{-{\bf p}}^l\delta(\omega + \omega_{{\bf p}}^l)$,
and finally function $I^l_p$ is defined by two-point correlation
function
\[
\langle A_{\mu}(p^{\prime})A_{\nu}(p)\rangle\sim \tilde{Q}_{\mu\nu}(p^{\prime})
I_{p^{\prime}}^l\delta(p^{\prime}-p).
\]
Taking into account above-mentioned, we derive the following expression
for estimation of the plasmon number density
\begin{equation}
N_{{\bf p}}^l\sim\omega_{{\bf p}}^l {\rm Z}_l^{-1}({\bf p})
\Bigl[\delta(\omega - \omega_{{\bf p}}^l)\Bigr]^{-1}
\Bigl[\delta(p^{\prime} - p)\Bigr]^{-1}
\langle A_{\mu}(p^{\prime})A_{\nu}(p)\rangle.
\label{eq:6e}
\end{equation}
Let us consider more typical two values of amplitude of a gauge field $A_{\mu}$
at the soft momentum scale . Using Eq.\,(\ref{eq:6e}) we have:
\begin{enumerate}
\item
${\rm let}\;\vert A_{\mu}(X)\vert\sim T\;(\vert A_{\mu}(p)\vert\sim
1/g(gT)^3),\;{\rm then}\;N_{\bf p}^l\,\sim\,\displaystyle\frac{1}{g^2},
\hspace{4.15cm}(6.4)$
\item
${\rm let}\;\vert A_{\mu}(X)\vert\sim\sqrt{g}T\;(\vert A_{\mu}(p)\vert\sim
1/\sqrt{g}(gT)^3),\;{\rm then}\;N_{\bf p}^l\,\sim\,\displaystyle\frac{1}{g}.
\hspace{3.4cm}(6.5)$
\end{enumerate}

For latter value of amplitude of the gauge field, when
$g\ll 1$, the occupation number $N_{\bf p}^l$ is large, and the use of
classical description is justified. The value of the
amplitude of a gauge field $(\sim \sqrt{g}T)$ is more probably at the momentum
and energy scale of the plasmon mode\footnote{Blaizot and Iancu \cite{bla1}
showed in the special case when the soft fields were thermal fluctuations at
soft scale $gT$, that their typical amplitudes would be of the order
$\vert A_{\mu} (X) \vert \sim \sqrt{g}T$. A similar estimation
of a value of the classical fields at the soft scale was shown by Nauta and
van Weert in Ref.\,\cite{nau} based upon a somewhat different context.}.
Power counting of the decay and regenerating rates, (\ref{eq:2r}) and
(\ref{eq:2t}) with regard to (\ref{eq:6q}) and (\ref{eq:6w})
gives the following estimation
\[
\Gamma_{\rm d,\,i}^{(2n+1)\,a a^{\prime}}\sim g^{4n}(gT)N_c^{2n}
(N_{\bf p}^l)^{2n+1}.
\]
If we now set $N_{\bf p}^l\,\sim\,\displaystyle\frac{1}{g^{\alpha}},\;
\alpha\!>\!0,$ then from the last expression it follows
$$
\Gamma_{\rm d,\,i}^{(2n+1)\,a a^{\prime}}\sim g^{2n(2-\alpha)}
g^{1-\alpha}N_c^{2n}T.
\eqno{(6.6)}
$$
For small value of oscillation amplitude (6.5) we have
\[
\Bigl(\Gamma_{\rm d,\,i}^{(2n+1)\,a a^{\prime}}\Bigr)_{A\sim\sqrt{g}T}
\sim g^{2n}N_c^{2n}T.
\]
From this estimation it can be seen that each subsequent term in the
functional expansions
(\ref{eq:2e}) is suppressed by more power of $g^2$ and here, we can only
restrict ourselves to first leading term, describing four-plasmon
decay process. Besides, here, the use of the linearized
Boltzmann equation for colorless plasmons, i.e. Eq.\,(\ref{eq:5o}) is justified ,
such the Planck distribution, relative of which the departure of the plasmon
number density is defined, is of order
\[
N_{\rm eq}^l({\bf p})\sim\displaystyle\frac{T}{\omega_{\bf p}^l}
\sim\displaystyle\frac{1}{g}.
\]
In this case it is said that a theory of a plasmon-plasmon interaction for
small amplitude of soft excitations is linear (linear amplitude regime),
and nonlinear effects connected with nonlinear terms over off-equilibrium
fluctuations of plasmon occupation number $\delta N_{\bf p}^l$  can
be treated as perturbations.

A situation is qualitatively changed, when a system is highly excited. In
a limiting case of a strong field, $A_{\mu}\sim T$, for $\alpha = 2$
from the estimation (6.6) it follows that functions
$\Gamma_{\rm d,\,i}^{(2n + 1) a a_1}$ is not to depend on
$n$. All terms in the expansions (\ref{eq:2e}) become of the same order
in magnitude, and the problem of resummation of all relevant contributions
arises. It is evident that a procedure of a linearization of a kinetic equation
for plasmon occupation number $N_{\bf p}^l$ in this case becomes unacceptable
and here, we lead to truly nonlinear interaction theory of the soft excitations
in hot QCD plasma.

It is noted an interesting analogy with phenomenon of so-called parton
saturation, which is of great interest in the physics of nuclear and hadronic processes
in the regime where Bjorken's $x$ becomes very small \cite{jal1, ian}. Saturation is
expected at gluon phase-space density of order $1/g^2$ (this is just maximum
value of the plasmon occupation number that in our case corresponds to maximum
value  of an oscillation amplitude of soft gluon fields:
$A_{\mu}\sim T$). Using this fact McLerran and
Venugopalan \cite{mcl} suggested a classical effective theory to describe
the gluon distribution in large nuclei and valid for some range of $x$.
The consequence of
a development of this theory is a construction by Jalilian-Marian, Kovner,
Leonidov and Weigert of renormalization group equation \cite{jal2}
(the JKLW equation) that predicts the evolution of the gluon distribution
function. In the saturation regime, where $A_{\mu}\sim{1/g}$, a
contributions of
all orders in the strong background fields which represent the condensate, should
be taken into account. In the low density, or weak-field limit, this equation
is linearized and reduced to known equations. In our case the role of the JKLW
equation fulfils a kinetic equation (\ref{eq:2w}) that is complete nonlinear
in a limiting case of a highly excitations of a soft-gluon field
$A_{\mu} \sim T$ and is reduced to a linearized Boltzmann equation
(\ref{eq:5o}) in the weak-field limit.\\

\section{\bf The matrix elements for $(2n + 2)$-plasmon decays}
\setcounter{equation}{0}

As shown in section 4, a calculation of matrix elements
${\rm M}^{a a_1 \ldots a_{2n + 1}}$ which are responsible for
processes of  $(2n + 2)$-plasmon decays reduces to
calculation of certain effective currents $\tilde{J}_{\mu}^{(2n + 1)}$
being functionals of $2n + 2$ free soft-gluon fields
$A_{\mu}^{(0)},\, n = 1, 2 \ldots$. The coefficient functions
in integrand of effective currents (\ref{eq:4o}) by equation (\ref{eq:4g})
define matrix elements of decay processes in tree approximation,
which allowed by
conservation laws of energy and momentum. In section 5 an example of construction
of such an effective current in simplest case of four-plasmon decay was
presented. However as we have shown in previous section, it becomes necessary
to take into account the contributions of higher decay processes,
i.e. subsequent terms in expansions (\ref{eq:2e}) when the energy level of
the soft plasma excitations increases. Here, we consider the problem of
determination of explicit form of matrix elements for higher decay processes
leaving aside more subtle questions connected with summing and convergence of
a series (\ref{eq:2e}).

As we have seen in section 5 the effective currents appear in the solution of
the nonlinear field equation (\ref{eq:3i}), that defines interacting
soft-gluon field $A_{\mu}$ in the form of a functional expansion in
free field $A_{\mu}^{(0)}$. Here, we shall extend this approach for determination
of the effective amplitudes $^{\ast}\tilde{\Gamma}^{(s)}$ for $s$ arbitrary values.
Let us rewrite here, the field equation (\ref{eq:3i}) for convenience of further
references
\begin{equation}
\,^{\ast}\tilde{\cal D}^{-1 \, \mu \nu}(p) A^{a}_{\nu}(p) =
-J^{a\mu}_{NL}[A]\equiv - \sum_{s=2}^{\infty} J^{(s)a\mu}(A,\ldots,A),
\label{eq:7q}
\end{equation}
where
\begin{equation}
J^{(s)a}_{\mu}(A,\ldots,A) =  \frac{1}{s!}\,g^{s-1}\!\!\int\!\!
\,^{\ast}\Gamma^{a a_1\ldots a_s}_{\mu\mu_1\ldots\mu_s}(p,-p_{1},\ldots,-p_{s})
A^{a_1\mu_1}(p_{1})A^{a_2\mu_2}(p_{2})\ldots A^{a_s\mu_s}(p_{s})
\label{eq:7w}
\end{equation}
\[
\times\delta (p - \sum_{i=1}^{s}p_{i})\prod_{i=1}^{s}dp_{i}.
\]

On the right-hand side of Eq.\,(\ref{eq:7w}) we discard all terms, containing
the thermal averaging (see discussion following by Eq.\,(\ref{eq:4o})
in section 4).
The nonlinear field equation (\ref{eq:7q}) can be formally solved perturbatively
(at least in the weak-field limit) order by order in the coupling
constant. We expand
\begin{equation}
A_{\mu}^a(p)=\sum_{s=1}^{\infty}A_{\mu}^{(s-1)\,a}(p),
\label{eq:7e}
\end{equation}
where $A_{\mu}^{(s-1)\,a}(p)$ is a contribution of order $g^{s-1}$ to the
soft-gluon field. For $s=1$ we have a solution for a free field. Substituting
an expansion (\ref{eq:7e}) into Eq.\,(\ref{eq:7q}) we obtain iterative solutions
of higher order in $s$:
\[
A_{\mu}^{(1)a}(p) =
-^{\ast}\tilde{\cal D}_{\mu\nu}(p)\tilde{J}^{(2)a\nu}(A^{(0)},A^{(0)}),
\]
\[
A_{\mu}^{(2)a}(p) =
-^{\ast}\tilde{\cal D}_{\mu\nu}(p)\tilde{J}^{(3)a\nu}(A^{(0)},A^{(0)},A^{(0)}),
\]
\begin{equation}
A_{\mu}^{(3)a}(p) = -^{\ast}\tilde{\cal D}_{\mu\nu}(p)
\tilde{J}^{(4)a\nu}(A^{(0)},A^{(0)},A^{(0)},A^{(0)}),
\label{eq:7r}
\end{equation}
\[
\ldots
\]
\[
A_{\mu}^{(s-1)a}(p) = -^{\ast}\tilde{\cal D}_{\mu\nu}(p)\tilde{J}^{(s)a\nu}
(A^{(0)},\ldots,A^{(0)}),
\]
\[
\ldots\;.
\]
Here, the effective current $\tilde{J}^{(3)a\nu}$ is defined by
Eqs.\,(\ref{eq:5r})\,--\,(\ref{eq:5y}), the current $\tilde{J}^{(4)a\nu}$
is defined by next iteration
\[
\tilde{J}^{(4)a\nu}(A^{(0)},A^{(0)},A^{(0)},A^{(0)})\equiv
J^{(4)a\nu}(A^{(0)},A^{(0)},A^{(0)},A^{(0)})
\]
\[
-\,\Bigl\{\,J^{(3)a\nu}(
-^{\ast}\tilde{\cal D}\tilde{J}^{(2)}(A^{(0)},A^{(0)}),A^{(0)},A^{(0)}) +
J^{(3)a\nu}(
A^{(0)},-^{\ast}\tilde{\cal D}\tilde{J}^{(2)}(A^{(0)},A^{(0)}),A^{(0)})
\]
\begin{equation}
+\,J^{(3)a\nu}(
A^{(0)},A^{(0)},-^{\ast}\tilde{\cal D}\tilde{J}^{(2)}(A^{(0)},A^{(0)}))\Bigr\}
\label{eq:7t}
\end{equation}
\[
-\,\Bigl\{\,J^{(2)a\nu}(
-^{\ast}\tilde{\cal D}\tilde{J}^{(3)}(A^{(0)},A^{(0)},A^{(0)}),A^{(0)})
+J^{(2)a\nu}(
A^{(0)},-^{\ast}\tilde{\cal D}\tilde{J}^{(3)}(A^{(0)},A^{(0)},A^{(0)}))\Bigr\}
\]
\[
-\,J^{(2)a\nu}(-^{\ast}\tilde{\cal D}\tilde{J}^{(2)}(A^{(0)},A^{(0)}),
-^{\ast}\tilde{\cal D}\tilde{J}^{(2)}(A^{(0)},A^{(0)}))
\]
etc. A similar iterative procedure of computing perturbative solutions
of the classical
Yang-Mills equation in various modifications is often used in solving
specific problems. From more closely works to subject of our
research mention may be made of the works of Kovchegov and Rischke \cite{kov},
Matinyan, M\"uller and Rischke \cite{mat} concerned with the problem of the
classical gluon radiation in ultrarelativistic nucleus-nucleus collisions.
Furthermore, in Ref.\,\cite{bod} B\"odeker used iterative solutions of the
classical Yang-Mills equations for soft modes of the gluon field in deriving
the Boltzmann equation for hard transverse gluons. Finally in the construction
of the correlation functions for ultrasoft fields
$(\omega,\,\vert {\bf p} \vert\sim g^2T)$, a similar iterative procedure was
used by Guerin in Ref.\,\cite{gue}.

However such a direct approach for determination of an explicit form of the
higher effective amplitudes $^{\ast}\tilde{\Gamma}^{(s)},\,s\!>\!3$, becomes
very complicated and as consequence ineffective.
In particular this associated with a necessity of additional symmetrization
of integrands, which becomes extremely intricate with increasing order $s$.
It is easily seen, for example, in attempt to determine the effective
amplitude $^{\ast}\tilde{\Gamma}^{(5)}$ immediately from the right-hand side of
Eq.\,(\ref{eq:7t}).
Here we proposed a somewhat different approach of calculation of
$^{\ast}\tilde{\Gamma}^{(s)}$. It is not transparent as the approach
based on iterations (\ref{eq:7r}), but it allows us to avoid many intermediate
operations and somewhat automate calculation procedure.

The calculating algorithm is based on a simple idea. The nonlinear current
$J_{NL}^{a \mu} [A]$ has two representations: by means of free and interacting
fields, which must be equal each other
\begin{equation}
J^{a\mu}_{NL}[A]=\sum_{s=2}^{\infty}\tilde{J}^{(s)a\mu}(A^{(0)},\ldots,A^{(0)})
=\sum_{s=2}^{\infty} J^{(s)a\mu}(A,\ldots,A).
\label{eq:7y}
\end{equation}
Here, the interacting fields on the right-hand side of the last equation are
defined by expansion
\begin{equation}
A^{a\mu}(p) = A^{(0)\,a\mu}(p) -\!
\,^{\ast}\tilde{\cal D}^{\mu\mu^{\prime}}(p)\sum_{s=2}^{\infty}
\tilde{J}^{(s)a}_{\mu^{\prime}}(A^{(0)},\ldots,A^{(0)}),
\label{eq:7u}
\end{equation}
where
\begin{equation}
\tilde{J}^{(s)a}_{\mu}(A^{(0)},\ldots,A^{(0)})
= \frac{1}{s!}\,g^{s-1}\!\!\int\!
\,^{\ast}\tilde{\Gamma}^{a a_1\ldots a_s}_{\mu\mu_1\ldots\mu_s}
(p,-p_{1},\ldots,-p_{s})
A^{(0)a_1\mu_1}(p_{1})\ldots A^{(0)a_s\mu_s}(p_{s})
\label{eq:7i}
\end{equation}
\[
\times\delta (p - \sum_{i=1}^{s}p_{i})\prod_{i=1}^{s}dp_{i}.
\]
Substitution of Eqs.\,(\ref{eq:7u}) and (\ref{eq:7i}) into (\ref{eq:7y})
turns this equation into identity. Now we functionally differentiate
left- and right-hand sides of equality (\ref{eq:7y}) with respect to
free field $A^{(0) a \mu}$ considering Eq.\,(\ref{eq:7i}) for differentiation
on the left-hand side and
Eqs.\,(\ref{eq:7w}), (\ref{eq:7u}) and (\ref{eq:7i}) for differentiation
on the right-hand side, and set $A^{(0) a \mu} = 0$
after all calculations. The required effective amplitudes
$^{\ast} \tilde{\Gamma}^{(s)}$ will appear both on the left-hand side
and on the right-hand side. However main effect is that the effective amplitudes
on the right-hand side will have at least one less external soft leg than on
the left-hand side. This enables us to calculate them in a recurrent way.
Below we shall give a few examples.

The second differentiation of a nonlinear current yields
\[
\left.\frac{\delta^2J_{NL\,\mu}^{a}[A](p)}
{\delta A^{(0)\,a_1\mu_1}(p_1)\delta A^{(0)\,a_2\mu_2}(p_2)}\,
\right|_{A^{(0)}=0}
=
\,^{\ast}\tilde{\Gamma}^{a a_1a_2}_{\mu \mu_1 \mu_2}(p,-p_{1},-p_{2})
\delta^{(4)}(p-p_1-p_2)
\]
\[
=\,^{\ast}\Gamma^{a a_1a_2}_{\mu \mu_1 \mu_2}(p,-p_{1},-p_{2})
\delta^{(4)}(p-p_1-p_2).
\]
This example shows that the effective amplitude of three-plasmon decay
coincide with usual three-gluon HTL-amplitude. This decay process is kinematically
forbidden and hence the $\delta$-function no support on the plasmon mass-shell.

The first interesting case arises in calculation of the next derivative.
It defines the effective amplitude for four-plasmon decay
\[
\left.\frac{\delta^3J_{NL\,\mu}^{a}[A](p)}
{\delta A^{(0)\,a_1\mu_1}(p_1)\delta A^{(0)\,a_2\mu_2}(p_2)
\delta A^{(0)\,a_3\mu_3}(p_3)}\,
\right|_{A^{(0)}=0}
\]
\[
=\,^{\ast}\tilde{\Gamma}^{a a_1a_2a_3}_{\mu\mu_1\mu_2\mu_3}
(p,-p_{1},-p_{2},-p_3)
\delta^{(4)}(p-p_1-p_2-p_3)
=\Bigl[\,^{\ast}\Gamma^{a a_1a_2a_3}_{\mu\mu_1\mu_2\mu_3}
(p,-p_{1},-p_{2},-p_3)
\]
\[
-\Bigl\{\Gamma^{aba_2}_{\mu\nu\mu_2}(p,-p+p_2,-p_2)
\,^{\ast}\tilde{\cal D}^{\nu\nu^{\prime}}(p_1+p_3)\!
\,^{\ast}\tilde{\Gamma}_{\nu^{\prime}\mu_1\mu_3}^{ba_1a_3}(p_1+p_3,-p_1,-p_3)
+({\rm circular\, perms.\,of}\,1,2,3)\Bigr\}\Bigr]
\]
\[
\times\delta^{(4)}(p-p_1-p_2-p_3).
\]
By using Jacobi identity for anti-symmetric structure constants, the right-hand
side of the last expression can be led to the expression (\ref{eq:5t})
obtained before. It can be viewed as an expansion of  an effective
amplitude $^{\ast}\tilde{\Gamma}_{\mu \mu_1 \mu_2 \mu_3}^{a a_1 a_2 a_3}$
in terms of two independent combinations of a product of two
structure constants
\begin{equation}
f^{a a_1 b} f^{b a_2 a_3}, \quad f^{a a_2 b} f^{b a_1 a_3},
\label{eq:7o}
\end{equation}
as it occurs for usual four-gluon HTL-amplitude \cite{bra1}. A similar
expansions take place also
for higher $^{\ast}\tilde{\Gamma}^{(s)}$ functions. Let us consider
the fourth derivative of a nonlinear current with respect to $A^{(0)a\mu}$,
defining a five-point effective amplitude. After rather cumbersome but direct
calculations we obtain
\begin{equation}
\left.\frac{\delta^4\!J_{NL\,\mu}^{a}[A](p)}
{\delta A^{(0)\,a_1\mu_1}(p_1)\ldots\delta A^{(0)\,a_4\mu_4}(p_4)}
\,\right|_{A^{(0)}=0}
\label{eq:7p}
\end{equation}
\[
=\,^{\ast}\tilde{\Gamma}^{a a_1\ldots a_4}_{\mu\mu_1\ldots\mu_4}
(p,-p_{1},\ldots,-p_{4})\delta^{(4)}(p-\sum_{i=1}^{4}p_i)
=\Bigl[
\,^{\ast}\Gamma^{a a_1\ldots a_4}_{\mu\mu_1\ldots\mu_4}
(p,-p_{1},\ldots,-p_{4})
\]
\[
-\,\Bigl\{\Gamma^{aba_2a_3}_{\mu\nu\mu_2\mu_3}(p,-p_1-p_4,-p_2,-p_3)
\,^{\ast}\tilde{\cal D}^{\nu\nu^{\prime}}(p_1+p_4)\!
\,^{\ast}\tilde{\Gamma}_{\nu^{\prime}\mu_1\mu_4}^{ba_1a_4}(p_1+p_4,-p_1,-p_4)
\]
\[
+\,\Gamma^{aa_1ba_3}_{\mu\mu_1\nu\mu_3}(p,-p_1,-p_2-p_4,-p_3)
\,^{\ast}\tilde{\cal D}^{\nu\nu^{\prime}}(p_2+p_4)\!
\,^{\ast}\tilde{\Gamma}_{\nu^{\prime}\mu_2\mu_4}^{ba_2a_4}(p_2+p_4,-p_2,-p_4)
\]
\[
+\,({\rm circular\, perms.\,of}\,2,3,4)\Bigr\}
\]
\[
-\Bigl\{\Gamma^{aba_1}_{\mu\nu\mu_1}(p,-p+p_1,-p_1)
\,^{\ast}\tilde{\cal D}^{\nu\nu^{\prime}}(p_2+p_3+p_4)\!
\,^{\ast}\tilde{\Gamma}_{\nu^{\prime}\mu_2\mu_3\mu_4}^{ba_2a_3a_4}(p-p_1,-p_2,-p_3,-p_4)
\]
\[
+\,({\rm circular\, perms.\,of}\,1,2,3,4)\Bigr\}
\]
\[
+\Bigl\{\Gamma^{abc}_{\mu\nu\lambda}(p,-p_1-p_3,-p_2-p_4)
\,^{\ast}\tilde{\cal D}^{\nu\nu^{\prime}}(p_1+p_3)\!
\,^{\ast}\tilde{\Gamma}_{\nu^{\prime}\mu_1\mu_3}^{ba_1a_3}(p_1+p_3,-p_1,-p_3)
\]
\[
\times\,^{\ast}\tilde{\cal D}^{\lambda\lambda^{\prime}}(p_2+p_4)\!
\,^{\ast}\tilde{\Gamma}_{\lambda^{\prime}\mu_2\mu_4}^{ca_2a_4}
(p_2+p_4,-p_2,-p_4)
+({\rm circular\, perms.\,of}\,2,3,4)\Bigr\}\Bigr]
\delta^{(4)}(p-\sum_{i=1}^{4}p_i).
\]
The obtained expression as it stands is very complicated and has involved
color structure. By direct algebraic transformations, using the Jacobi identity
for structure constants and relations (\ref{eq:5s}) and (\ref{eq:5d})
it can be represented in the form of an expansion in terms of six independent
combinations of a product of three structure constants, which we are choosen
as follows
\begin{equation}
f^{a a_1 b_1} f^{b_1 a_2 b_2} f^{b_2 a_3 a_4}, \quad
f^{a a_1 b_1} f^{b_1 a_3 b_2} f^{b_2 a_2 a_4}, \quad
f^{a a_3 b_1} f^{b_1 a_2 b_2} f^{b_2 a_1 a_4},
\label{eq:7a}
\end{equation}
\[
f^{a a_2 b_1} f^{b_1 a_3 b_2} f^{b_2 a_1 a_4}, \quad
f^{a a_3 b_1} f^{b_1 a_1 b_2} f^{b_2 a_2 a_4}, \quad
f^{a a_2 b_1} f^{b_1 a_1 b_2} f^{b_2 a_3 a_4}.
\]
Here, as in basis (\ref{eq:7o}), we fixed the first and the last color indices.
The expansion $^{\ast}\tilde{\Gamma}^{(5)}$ in terms of a basis (\ref{eq:7a})
reads
\begin{equation}
\,^{\ast}\tilde{\Gamma}^{a a_1\ldots a_4}_{\mu\mu_1\ldots\mu_4}
(p,-p_{1},\ldots,-p_{4})=
f^{a a_1 b_1} f^{b_1 a_2 b_2} f^{b_2 a_3 a_4}
\,^{\ast}\tilde{\Gamma}_{\mu\mu_1\ldots\mu_4}
(p,-p_{1},\ldots,-p_{4})
\label{eq:7s}
\end{equation}
\[
+\,({\rm perms.\,of}\,1,2,3),
\]
where a sum is over all $3!$ permutations of the external gluons 1, 2 and 3,
and effective subamplitude is defined as
\[
\,^{\ast}\tilde{\Gamma}_{\mu\mu_1\ldots\mu_4}(p,-p_{1},\ldots,-p_{4})
\equiv\,^{\ast}\Gamma_{\mu\mu_1\ldots\mu_4}(p,-p_{1},\ldots,-p_{4})
\]
\[
-\,\Bigl\{
\Gamma_{\mu\mu_1\mu_2\nu}(p,-p_1,-p_2,-p_3-p_4)
\,^{\ast}\tilde{\cal D}^{\nu\nu^{\prime}}(p_3+p_4)\!
\,^{\ast}\tilde{\Gamma}_{\nu^{\prime}\mu_3\mu_4}(p_3+p_4,-p_3,-p_4)
\]
\[
+\,({\rm circular\,perms.\,of}\,2,3,4)\Big\}
\]
\begin{equation}
-\Bigl\{\Gamma_{\mu\mu_1\nu}(p,-p_1,-p+p_1)
\,^{\ast}\tilde{\cal D}^{\nu\nu^{\prime}}(p-p_1)\!
\,^{\ast}\tilde{\Gamma}_{\nu^{\prime}\mu_2\mu_3\mu_4}(p-p_1,-p_2,-p_3,-p_4)
\label{eq:7d}
\end{equation}
\[
+\,\Gamma_{\mu\nu\mu_4}(p,-p+p_4,-p_4)
\,^{\ast}\tilde{\cal D}^{\nu\nu^{\prime}}(p-p_4)\!
\,^{\ast}\tilde{\Gamma}_{\nu^{\prime}\mu_1\mu_2\mu_3}(p-p_4,-p_1,-p_2,-p_3)
\Bigr\}
\]
\[
+\,\Gamma_{\mu\nu\lambda}(p,-p_1-p_2,-p_3-p_4)
\,^{\ast}\tilde{\cal D}^{\nu\nu^{\prime}}(p_1+p_2)\!
\,^{\ast}\tilde{\Gamma}_{\nu^{\prime}\mu_1\mu_2}(p_1+p_2,-p_1,-p_2)
\]
\[
\times\,^{\ast}\tilde{\cal D}^{\lambda\lambda^{\prime}}(p_3+p_4)\!
\,^{\ast}\tilde{\Gamma}_{\lambda^{\prime}\mu_3\mu_4}(p_3+p_4,-p_3,-p_4).
\]

The examples considered above suggest that a color structure of
the effective amplitudes $^{\ast} \tilde{\Gamma}^{(s)}$ entirely coincides
with color structure of usual s-gluon HTL-amplitudes\footnote{
It is clear that if this statement is really true (for the present we
cannot rigorously prove this fact) then the algorithm of calculation
of $^{\ast} \tilde{\Gamma}^{(s)}$ is not optimal. It takes its additional
modification, which enables one to derive at once the effective
subamplitudes instead of the total effective amplitudes.}, derived by
Braaten and Pisarski \cite{bra3} for an arbitrary number of external
soft-gluon legs.
Besides, by direct calculation it can be shown that an effective subamplitude
(\ref{eq:7d}) satisfies two relations which hold for usual five-gluon
HTL-amplitude, namely
\begin{equation}
\,^{\ast}\tilde{\Gamma}_{\mu\mu_1\mu_2\mu_3\mu_4}(p,-p_1,-p_2,-p_3,-p_4)\,+
\,^{\ast}\tilde{\Gamma}_{\mu\mu_3\mu_1\mu_2\mu_4}(p,-p_3,-p_1,-p_2,-p_4)
\label{eq:7f}
\end{equation}
\[
+\,^{\ast}\tilde{\Gamma}_{\mu\mu_1\mu_3\mu_2\mu_4}(p,-p_1,-p_3,-p_2,-p_4)
+\,^{\ast}\tilde{\Gamma}_{\mu\mu_1\mu_2\mu_4\mu_3}(p,-p_1,-p_2,-p_4,-p_3)=0,
\]
\[
\,^{\ast}\tilde{\Gamma}_{\mu\mu_1\mu_2\mu_3\mu_4}(p,-p_1,-p_2,-p_3,-p_4)=
-\,^{\ast}\tilde{\Gamma}_{\mu\mu_4\mu_3\mu_2\mu_1}(p,-p_4,-p_3,-p_2,-p_1).
\]
Note that the order of the spacetime indices in relations (\ref{eq:7f})
(as well as in relations (\ref{eq:5s}) and (\ref{eq:5d})) is important.
Unfortunately,
for the present we cannnot prove similar properties for an arbitrary effective
subamplitude $^{\ast} \tilde{\Gamma}^{(s)}$, since we have not in hand general
analytic expression for $^{\ast} \tilde{\Gamma}^{(s)}$. However relations
(\ref{eq:5s}), (\ref{eq:5d}), and (\ref{eq:7f})
for the first nontrivial $s = 4,\,5$ provide a reason for assumption that they
hold for a general case in a form, suggested by Braaten and Pisarski
(Eqs.\,(\ref{eq:2e}) and (\ref{eq:2i}) in Ref.\,\cite{bra3}) for an arbitrary
s-gluon HTL-amplitude.

However as was mentioned at the end of section 5 the Ward identities for
effective
amplitudes have not take place in the form as for usual HTL-amplitudes.
The relations
between higher and lower effective amplitudes can be obtained by direct
contraction of the expression (\ref{eq:5y}) and (\ref{eq:7d}) type with
momentum variable $p$. But it is more convenient to derive similar relations
in approach suggested by Blaizot and Iancu in Ref.\,\cite{bla2} through
corresponding differentiation of the conservation law for the color current,
which in the momentum representation has a form
\begin{equation}
p^{\mu}J_{\mu}^a(p) = -igf^{abc}\!\int\!A^{b\mu}(p_1)
J_{\mu}^c(p_2)\delta(p-p_1-p_2)dp_1dp_2.
\label{eq:7g}
\end{equation}
A gauge field $A^{b\mu}(p_1)$ on the right-hand side of Eq.\,(\ref{eq:7g}) is
considered as
interacting one, i.e. specified by a series (\ref{eq:7u}). It is convenient
to represent the color current $J^a_{\mu}(p)$ entering into Eq.\,(\ref{eq:7g})
as a sum of two parts: linear and nonlinear ones with respect to interacting
field $A_{\mu}^a(p)$
\begin{equation}
J^{a\mu}(p)=J^{a\mu}_{L}(p)\,+\,J^{a\mu}_{NL}(p).
\label{eq:7h}
\end{equation}
By relation (\ref{eq:7y}) a nonlinear part of the current can be represented
in the form of expansion in terms of a free field
\begin{equation}
J^{a\mu}_{NL}[A](p)\rightarrow
\tilde{J}^{a\mu}_{NL}[A^{(0)}](p)
=\sum_{s=2}^{\infty}\tilde{J}^{(s)a\mu}(A^{(0)},\ldots,A^{(0)}).
\label{eq:7j}
\end{equation}
By virtue of the fact that a current $J^{a\mu}_{NL}[A](p)$ involves
also bare three- and four-gluon vertices into its definition,
for consistency one need to take a linear part of the current in the form
\begin{equation}
J_{L}^{a\mu}(p) = -^{\ast}\tilde{\cal D}^{\mu\nu}(p)A_{\nu}^a(p),
\label{eq:7k}
\end{equation}
i.e. a coefficient of proportionality between $J_{L}^{a\mu}$ and $A^{a\mu}$
is here the inverse resumm gluon propagator (without gauge fixing term),
and not the polarization tensor, as in usual definition. Substituting
(\ref{eq:7h}) into (\ref{eq:7g}) with regard to (\ref{eq:7j}) and (\ref{eq:7k})
we result in the following expression instead of (\ref{eq:7g})
\begin{equation}
p_{\mu}\tilde{J}_{NL}^{a\mu}[A^{(0)}](p)
= -igf^{abc}\int\!A^{b}_{\mu}(p_1)
^{\ast}\tilde{\cal D}^{\mu\nu}(p_2)A_{\nu}^c(p_2)\delta(p-p_1-p_2)dp_1dp_2
\label{eq:7l}
\end{equation}
\[
-\,igf^{abc}\int\!A^{b}_{\mu}(p_1)
\tilde{J}_{NL}^{c\mu}[A^{(0)}](p_2)\delta(p-p_1-p_2)dp_1dp_2.
\]
Here, on the left-hand side we taken into account orthogonality of an inverse
propagator $p_{\mu}\!\,^{\ast}\tilde{{\cal D}}^{-1\,\mu\nu}(p) = 0$.
Differentiating
Eq.\,(\ref{eq:7l}) with respect to free field $A_{\mu}^{(0) a}$, considering
(\ref{eq:7u}) and (\ref{eq:7i}), one obtains relations between effective
amplitudes $^{\ast}\tilde{\Gamma}^{(s)}$. A contraction momentum
$p$ with the effective amplitude $^{\ast}\tilde{\Gamma}^{(s)}$ is expressed
here in the form of combination of the same effective amplitudes of lower
order, $^{\ast}\tilde{\Gamma}^{(s^{\prime})},\, s^{\prime}<s$.\\

\section{\bf Gauge invariance of the effective amplitudes}
\setcounter{equation}{0}

The properties (\ref{eq:5s}), (\ref{eq:5d}) and (\ref{eq:7f}) for effective
amplitudes are trivial in the sense that they separately hold for
every  typical group of terms, formed effective amplitudes\footnote{In the case
of four-plasmon function (\ref{eq:5y}) we look at two groups of terms
differ in structure: four-gluon HTL-amplitude and group consisting of
two terms with three-gluon HTL-amplitude. For five-point function (\ref{eq:7d})
we have already four groups of terms with essentially different structure.}.
These properties do not reflect more deep
connection between different terms involved into different groups. Here, we
consider one further property of the effective amplitudes, for which such
a connection completely manifests.

In proving the gauge-invariance of the nonlinear Landau damping rate,
we have shown \cite{mar1} that the function (\ref{eq:5a}) entering into the definition
of the matrix element (\ref{eq:5u}) can be introduced in its simplest
reduced form,
\[
\,^{\ast}\tilde{\Gamma} (p,-p_1,-p_2,-p_3)=p^2p_1^2p_2^2p_3^2\Bigl\{
\,^{\ast}{\Gamma}_{0000}(p,-p_{1},-p_{2},-p_3)
\]
\begin{equation}
-\,\Gamma_{00\nu}(p,-p_1,-p+p_1)
\,^{\ast}\tilde{\cal D}^{\nu\nu^{\prime}}(p_2+p_3)
\,^{\ast}\Gamma_{\nu^{\prime}00}(p_2+p_3,-p_2,-p_3)
\label{eq:8q}
\end{equation}
\[
\left.- \,\Gamma_{00\nu}(p,-p_3,-p+p_3)
\,^{\ast}\tilde{\cal D}^{\nu\nu^{\prime}}(p_1+p_2)
\,^{\ast}\Gamma_{\nu^{\prime}00}(p_1+p_2,-p_2,-p_1)
\Bigr\}\right|_{\rm on-shell}.
\]
The proof of validity of this reduction is based on the use of the Ward effective
identities for HTL-amplitudes and the plasmon mass-shell condition. The
right-hand side of last expression, as shown in Ref.\,\cite{mar1} is
identical both in temporal and covariant gauges. The latter is obtained
from Eq.\,(\ref{eq:5a}) by the following replacements of projector and
propagator, (see Eqs.\,(\ref{eq:3p})\,--\,(\ref{eq:3s}))
\[
\tilde{u}_{\mu}(p)\rightarrow\bar{u}_{\mu}(p),
\]
\begin{equation}
\,^{\ast}\tilde{\cal D}_{\mu\nu}(p) \rightarrow
\,^{\ast}{\cal D}_{\mu\nu}(p) =
- P_{\mu \nu}(p) \,^{\ast}\!\Delta^t(p) -
Q_{\mu \nu}(p) \,^{\ast}\!\Delta^l(p)
+\,\xi D_{\mu\nu}(p)\Delta^{0}(p),
\label{eq:8w}
\end{equation}
where
\[
P_{\mu \nu} (p) = g_{\mu \nu} -
D_{\mu \nu}(p) -
Q_{\mu \nu}(p), \quad
Q_{\mu \nu}(p) =
\frac{\bar{u}_{\mu} (p) \bar{u}_{\nu} (p)}{\bar{u}^2(p)},\quad
\Delta^0(p)=\frac{1}{p^2},
\]
and $\xi$ is a gauge parameter in the covariant gauge. All terms on the
right-hand side of Eq.\,(\ref{eq:8q}), which include gauge parameter, vanish on
mass-shell.

We can assume that a similar reduction holds for an arbitrary effective amplitude.
However a proof of this statement is impossible in the general case by reason
of absence of general expression for $^{\ast}\tilde{\Gamma}^{(s)}$, when
a proof by induction is allowed. The only thing that we can make is to consider
a contraction similar to (\ref{eq:8q}) for five-point effective amplitude,
exact form of which (\ref{eq:7d}) is known. It should be noted that as distinct
from (\ref{eq:8q}) given contraction is of no any physical meaning on the
plasmon mass-shell, because the process of five-plasmon decay is kinematically
forbidden. Since here, the intermediate calculations are cumbersome, we give only
the net result. Using effective Ward identities for HTL-amplitudes
\cite{bra1, bra3} and mass-shell condition, we derive
\[
\,^{\ast}\tilde{\Gamma}_{\mu\mu_1\ldots\mu_4}(p,-p_{1},\ldots,-p_{4})
\left.\tilde{u}^{\mu}(p)\tilde{u}^{{\mu}_1}(p_1)\ldots\tilde{u}^{{\mu}_4}(p_4)
\right|_{\rm on-shell}
= -p^2p_1^2p_2^2p_3^2p_4^2
\,\Bigl[
\]
\[
\,^{\ast}\Gamma_{00\ldots 0}(p,-p_{1},\ldots,-p_{4})
-\Bigl\{\,\Gamma_{000\nu}(p,-p_1,-p_2,-p_3-p_4)
\,^{\ast}\tilde{\cal D}^{\nu\nu^{\prime}}(p_3+p_4)\!
\,^{\ast}\tilde{\Gamma}_{\nu^{\prime}00}(p_3+p_4,-p_3,-p_4)
\]
\[
+\,({\rm circular\,perms.\,of}\,2,3,4)\Big\}
\]
\begin{equation}
-\,\Bigl\{\Gamma_{00\nu}(p,-p_1,-p+p_1)
\,^{\ast}\tilde{\cal D}^{\nu\nu^{\prime}}(p - p_1)\!
\,^{\ast}\tilde{\Gamma}_{\nu^{\prime}000}(p-p_1,-p_2,-p_3,-p_4)
\label{eq:8e}
\end{equation}
\[
+\,\Gamma_{0\nu0}(p,-p+p_4,-p_4)
\,^{\ast}\tilde{\cal D}^{\nu\nu^{\prime}}(p - p_4)\!
\,^{\ast}\tilde{\Gamma}_{\nu^{\prime}000}(p-p_4,-p_1,-p_2,-p_3)
\Bigr\}
\]
\[
+\,\Gamma_{0\nu\lambda}(p,-p_1-p_2,-p_3-p_4)
\,^{\ast}\tilde{\cal D}^{\nu\nu^{\prime}}(p_1+p_2)\!
\,^{\ast}\tilde{\Gamma}_{\nu^{\prime}00}(p_1+p_2,-p_1,-p_2)
\]
\[
\times\,^{\ast}\tilde{\cal D}^{\lambda\lambda^{\prime}}(p_3+p_4)\!
\,^{\ast}\tilde{\Gamma}_{\lambda^{\prime}00}(p_3+p_4,-p_3,-p_4)
\Bigr]_{\rm on-shell}.
\]
As we see, the result is similar to (\ref{eq:8q}). It is also easily to check
that all terms with a gauge parameter in Eq.\,(\ref{eq:8e}) vanish on mass-shell.
If we perform a replacements (\ref{eq:8w}) on the left-hand side
(i.e. define a matrix element of five-plasmon decay in a covariant gauge), then
after
analogous computations we lead to the same expression on the right-hand side
of (\ref{eq:8e}) with only distinction in a common sign\footnote{ Such distinct in
a sign in calculation of a similar contraction with an effective amplitude with
odd number of external legs can be already found for first three-point effective
amplitude $^{\ast}\tilde{\Gamma}_{\mu\mu_1\mu_2} (p,- p_1,-p_2)$.}. This fact of
a gauge non-invariance immediately points to the fact that odd effective
amplitudes taken on mass-shell of plasma excitations are of no physical meaning,
as said above. All decay processes involving odd numbers of plasmons are
kinematically forbidden. Therefore in an expansion of the nonlinear current
(\ref{eq:7j}) all effective currents $^{\ast}\tilde{J}_{\mu}^{(s)a}$
with even number of free fields $A^{(0)}_{\mu}$ (correspondingly containing
odd effective amplitudes) by $\delta$-functions are equal to zero on mass-shell.

At the end of this section we note that in spite of the fact that result
(\ref{eq:8e}) is of only pure methodological meaning, nevertheless two
examples (\ref{eq:8q}) and (\ref{eq:8e}) provide a reason to use of
considerably simple expressions
of (\ref{eq:8q}) type for all $(2n + 2)$-matrix elements
${\rm M}^{a a_1\ldots a_{2n+1}}$ in particular calculations.\\

\section{\bf The Vlasov-Boltzmann equation for color plasmons}
\setcounter{equation}{0}

In a previous sections we have considered the problem of deriving
kinetic equation of a Boltzmann type, describing decay processes involving
colorless plasmons. The number density of plasmons here has a trivial color
structure: $N_{\bf p}^{l\,ab} = \delta^{ab} N_{\bf p}^l$. Such a structure
of the plasmon number density takes place, when there is no external color
current or/and color mean field in the system. Now we assume that
a time-space dependent external perturbation (e.g. external color current
$j_{\mu}^{{\rm ext}\,a}(x)$) starts acting on the system.
In the presence of the
external color perturbation, the soft gauge field develops an expectation
value $\langle A_{\mu}^{a}(x)\rangle\equiv{\cal A}_{\mu}^a(x)\ne 0$,
and the number
density of the plasmons acquires a non-diagonal color structure. Thereby,
we lead to the problem of construction of more complicated
kinetic theory for color plasmons. The kinetic (matrix) equation in this
theory will take into account such purely non-Abelian effect as the precession
of the color charge of the plasmon, the existence of which qualitatively
differs non-Abelian plasma from Abelian one, where the plasmons do not
carry electric charge. Our further consideration will be to a certain extent
of phenomenological character, being only a first step towards to construction
of a kinetic theory for color plasmons in hot QCD plasma. For rigorous
justification of all assumption it should be invoked the methods of an
off-equilibrium field theory.

The assumption that a structure of the kinetic equation for color plasmons
is similar to the structure of the kinetic equation for hard transverse
gluons in the form proposed by Arnold, Son and Yaffe in Ref.\,\cite{arn},
will be our initial theoretical premise to construction of above-mentioned
transport theory. More precisely, we expect the time-space evolution of
$N_{\bf p}^l = (N_{\bf p}^{l\,ab})$ to be described by
\begin{equation}
\Bigl({\cal D}_t + {\bf V}_{\bf p}^l\cdot{\cal D}_{\bf x}\Bigr) N_{\bf p}^l
-\frac{1}{2}\,\Bigl\{\Bigl({\bf {\cal E}}(x) + ({\bf V}_{\bf p}^l\times
{\bf {\cal B}}(x))\Bigr)_i,\nabla_{{\rm p}_i}N_{\bf p}^l\Bigr\} =
-\,{\rm C}\,[N_{\bf p}^l],
\label{eq:9q}
\end{equation}
where ${\cal D}_{\mu}$ is a covariant derivative acting as
\[
{\cal D}_{\mu}N_{\bf p}^l \equiv \partial_{\mu}
+ig[{\cal A}_{\mu}(x),N_{\bf p}^l],
\]
${\cal A}_{\mu}=T^a{\cal A}_{\mu}^a$ is mean soft-gluon field expressed
in terms of Hermitian generators in adjoint representation
$T^a$ $((T^a)^{bc}=-i f^{abc},\;{\rm tr}(T^aT^b)=N_c\delta^{ab})$,
with $[\,,]$ and $\{\,,\}$
denoting the commutator and anticommutator in color space, respectively;
${\cal E}^i(x)$ and ${\cal B}^i(x)$ are the mean chromoelectric and
chromomagnetic fields.
Futhermore we consider that the collision term ${\rm C}\,[N_{\bf p}^l]$ has
a following structure \cite{arn}
\begin{equation}
{\rm C}\,[N_{\bf p}^l]=\frac{1}{2}\,\Bigl\{
N_{\bf p}^l,\Gamma_{\rm d}[N_{\bf p}^l]\Bigr\} -
\frac{1}{2}\,\Bigl\{( 1 + N_{\bf p}^l ),\Gamma_{\rm i}[N_{\bf p}^l]\Bigr\}
-\ldots,
\label{eq:9w}
\end{equation}
where $\Gamma_{\rm d}[N_{\bf p}^l]=(\Gamma_{\rm d}^{a a^{\prime}}[N_{\bf p}^l])$
and $\Gamma_{\rm i}[N_{\bf p}^l]=(\Gamma_{\rm i}^{a a^{\prime}}[N_{\bf p}^l])$
represent a generalized decay and regenerating rates of color plasmons,
respectively. These rates can be formally represented, like a colorless case
in the form of functional expansion in
powers of the number density of color plasmons
\begin{equation}
\Gamma_{\rm d}^{a a^{\prime}}[N_{\bf p}^l] = \sum_{n = 1}^{\infty}
\Gamma_{\rm d}^{(2n + 1)\,a a^{\prime}} [N_{\bf p}^l] , \; \;
\Gamma_{\rm i}^{a a^{\prime}} [N_{\bf p}^l] = \sum_{n = 1}^{\infty}
\Gamma_{\rm i}^{(2n + 1)\,a a^{\prime}} [N_{\bf p}^l],
\label{eq:9e}
\end{equation}
where
\begin{equation}
\Gamma_{\rm d}^{(2n + 1)\,a a^{\prime}} [N_{\bf p}^l]=
\int\! d{\cal T}^{(2n + 1)}\,{\it w}^{\{a^{\prime};\,a\}}_{2n + 2}
({\bf p}, {\bf p}_1, \ldots ,{\bf p}_n;
{\bf p}_{n + 1}, \ldots , {\bf p}_{2n + 1})
\, N_{{\bf p}_1}^{l\,a_1^{\prime} a_1}\ldots
N_{{\bf p}_n}^{l\,a_n^{\prime} a_n}\hspace{1.55cm}
\label{eq:9r}
\end{equation}
\[
\times(1 + N_{{\bf p}_{n + 1}}^{l})^{a_{n+1}a^{\prime}_{n+1}} \ldots
(1 + N_{{\bf p}_{2n + 1}}^{l})^{a_{2n+1}a^{\prime}_{2n+1}},
\]
\[
\Gamma_{\rm i}^{(2n + 1)\,a a^{\prime}} [N_{\bf p}^l]\!=\!\!
\int\!\! d{\cal T}^{(2n + 1)}
{\it w}^{\{a^{\prime};\,a\}}_{2n + 2}({\bf p}, {\bf p}_1, \ldots ,{\bf p}_n;
{\bf p}_{n + 1}, \ldots , {\bf p}_{2n + 1})
\,(1 +  N_{{\bf p}_1}^l)^{a_1^{\prime} a_1}\!\ldots\!
(1 + N_{{\bf p}_n}^l)^{a_n^{\prime} a_n}
\]
\begin{equation}
\times N_{{\bf p}_{n + 1}}^{l\,a_{n+1}a^{\prime}_{n+1}}
\ldots N_{{\bf p}_{2n + 1}}^{l\,a_{2n+1}a^{\prime}_{2n+1}} .
\label{eq:9t}
\end{equation}
Here, ${\it w}^{\{a^{\prime};\,a\}}_{2n + 2}
\equiv {\it w}^{a^{\prime}a_1^{\prime}\ldots
a^{\prime}_{2n+1};\,aa_1\ldots a_{2n + 1}}
({\bf p},{\bf p}_1,\ldots,{\bf p}_n;{\bf p}_{n + 1},\ldots,{\bf p}_{2n + 1})$
is the scattering probability defined as
\[
{\it w}^{\{a^{\prime};\,a\}}_{2n + 2}({\bf p}, {\bf p}_1, \ldots ,{\bf p}_n;
{\bf p}_{n + 1}, \ldots , {\bf p}_{2n + 1})=
\]
\[
= {\rm M}^{\ast \{a^{\prime}\}}({\bf p},{\bf p}_1,\ldots,
{\bf p}_n,-{\bf p}_{n+1},\ldots, -{\bf p}_{2n+1})
{\rm M}^{\{a\}}({\bf p},{\bf p}_1,\ldots,{\bf p}_n,-{\bf p}_{n+1},\ldots,
-{\bf p}_{2n+1})
\]
\[
+\sum_{1\leq i_1\leq n}^{({\bf p}_{i_1}a_{i_1})}
{\rm M}^{\ast\{a^{\prime}\}}{\rm M}^{\{a\}}
+\sum_{1\leq i_1<i_2\leq n}^{({\bf p}_{i_1}a_{i_1},{\bf p}_{i_2}a_{i_2})}
{\rm M}^{\ast\{a^{\prime}\}}{\rm M}^{\{a\}}+
\ldots+\sum^{({\bf p}_1a_1,\ldots,{\bf p}_na_n)}
{\rm M}^{\ast\{a^{\prime}\}}{\rm M}^{\{a\}}
\]
with {\it the same} matrix elements ${\rm M}^{\{a\}} ({\bf p},{\bf p}_{1},\ldots
{\bf p}_{n},-{\bf p}_{n + 1},\ldots,-{\bf p}_{2n + 1})$ as was
defined by us in colorless case by general relation (\ref{eq:4g})
(see, however the end of this section). Here, the summing symbol
$\sum_{1\leq i_1 \leq n}^{({\bf p}_{i_1}a_{i_1})}$
designates that besides summing over all possible interchange of momenta
$p_{i_1} \leftrightarrow - p_{j_1}$, where indicies $i_1,\,j_1$ run
$1\leq i_1 \leq n$, $n+1\leq j_1 \leq 2n +1,$ it needs to interchange and
color indices $a_{i_1} \leftrightarrow  a_{j_1}$,
$a_{i_1}^{\prime} \leftrightarrow  a_{j_1}^{\prime}$ at one time.
The other summing symbols are taken in a similar fashion.
In arrangement of color indices in Eqs.\,(\ref{eq:9r}) and
(\ref{eq:9t}) we follow rule proposed by Arnold, Son and Yaffe in
Ref.\,\cite{arn}
in the context of deriving collision term for hard color particles. The dots on
the righ-hand side of Eq.\,(\ref{eq:9w}) is refered to contributions containing
commutators of $[\,{\rm Re}\,\Pi_{R}, N_{\bf p}^l]$ type, where $\Pi_{R}$
is retarded soft-gluon self-energy.

In the case of four-plasmon decay, the scattering probability
${\it w}_4^{\{a^{\prime};\,a\}}$ by general expression has a form
\[
{\it w}_4^{\{a^{\prime};\,a\}}({\bf p},{\bf p}_1;{\bf p}_2,{\bf p}_3)=
M^{\ast a^{\prime} a_1^{\prime} a_2^{\prime} a_3^{\prime}}
({\bf p}, {\bf p}_1, - {\bf p}_2, - {\bf p}_3)
M^{a a_1 a_2 a_3}({\bf p}, {\bf p}_1, - {\bf p}_2, - {\bf p}_3)
\]
\[
+\,M^{\ast a^{\prime} a_2^{\prime} a_1^{\prime} a_3^{\prime}}
({\bf p}, - {\bf p}_2, {\bf p}_1, - {\bf p}_3)
M^{a a_2 a_1 a_3}({\bf p}, - {\bf p}_2, {\bf p}_1, - {\bf p}_3)
\]
\[
+\,M^{\ast a^{\prime} a_3^{\prime} a_2^{\prime} a_1^{\prime}}
({\bf p}, - {\bf p}_3, - {\bf p}_2,  {\bf p}_1)
M^{a a_3 a_2 a_1}({\bf p}, - {\bf p}_3, - {\bf p}_2, {\bf p}_1).
\]
By direct calculation, using an explicit expression (\ref{eq:5t}) for
color structure of an effective amplitude
$^{\ast} \tilde{\Gamma}_4^{\{a\}\{\mu\}}$ and
properties for subamplitude $^{\ast}\tilde{\Gamma}_4^{\{\mu\}}$
(\ref{eq:5s}), (\ref{eq:5d}),
it can be shown  that a sum of two last terms in this expression is equal to
doubled value of a first term.

It needs to supplement the Vlasov-Boltzmann equation (\ref{eq:9q}) by
mean field equation, defining a change of mean field in a system
in self-consistent manner
\[
{\cal D}^{\nu}(x){\cal F}_{\mu\nu}(x) = j_{\mu}^{\rm plasm}(x)
+ j_{\mu}^{\rm ext}(x),
\]
where induced current $j_{\mu}^{\rm plasm}(x)=(j_{0}^{\rm plasm}(x),\,
{\bf j}^{\rm plasm}(x)),$
\begin{equation}
j_{0}^{\rm plasm}(x)= gT^a\!\!\int\!\frac{d{\bf p}}{(2\pi)^3}\,
{\rm tr}(T^aN_{\bf p}^l),
\quad
{\bf j}^{\rm plasm}(x)= gT^a\!\!\int\!\frac{d{\bf p}}{(2\pi)^3}\,
\frac{\partial\omega_{\bf p}^l}{\partial{\bf p}}\,{\rm tr}(T^aN_{\bf p}^l)
\label{eq:9y}
\end{equation}
is the color current resulting from color-plasmon number density and
$j_{\mu}^{\rm ext}$ is the external current plays a part of initial color
perturbation.

It is clear that dynamics of nonlinear interaction of soft color excitations
is considerably more intricate as compared with colorless ones.
To form a certain notion of this fact we consider in detail the
linearized version of Vlasov-Boltzmann equation (\ref{eq:9q}), keeping only
the first term, $n = 1$, in expansions (\ref{eq:9e}). We write the number
density of color plasmons as
\[
N^{l\,ab}_{\bf p} = N_{eq}^l({\bf p})\delta^{ab} + \delta N^{l\,ab}_{\bf p},
\]
and parametrizate the off-equilibrium fluctuation of the occupation number
similar to Eq.\,(\ref{eq:5i}), where now the function ${\cal W}^l_{\bf p}$
is a color matrix in the adjoint representation:
${\cal W}^l_{\bf p}={\cal W}^{l\,a}_{\bf p}T^a$. After some cumbersome
algebraic transformations, we derive a linearized kinetic equation for the
color plasmons from Eqs.\,(\ref{eq:9q})\,--\,(\ref{eq:9t})
\begin{equation}
\Bigl({\cal D}_t + {\bf V}_{\bf p}^l\cdot{\cal D}_{\bf x}\Bigr)
{\cal W}_{\bf p}^l=
-\,g\,({\bf V}_{\bf p}^l\cdot{\bf {\cal E}}(x))
-\,\delta{\rm C}\,[{\cal W}_{\bf p}^l],
\label{eq:9u}
\end{equation}
where linearized collision term $\delta{\rm C}$ has a following form
\[
\,\delta{\rm C}\,[{\cal W}_{\bf p}^l] =
\int\!
\frac{d{\bf p}_1}{(2\pi)^3}\frac{d{\bf p}_2}{(2 \pi)^3}
\frac{d{\bf p}_3}{(2 \pi)^3} \,
(2 \pi)^4\delta (\omega_{\bf p}^l+\omega_{{\bf p}_1}^l-\omega_{{\bf p}_2}^l -
\omega_{{\bf p}_3}^l)
\]
\begin{equation}
\times\!\delta({\bf p}+{\bf p}_1 - {\bf p}_2 - {\bf p}_3)\,
\frac{N_{eq}^l({\bf p}_1)(N_{eq}^l({\bf p}_2) + 1)
(N_{eq}^l({\bf p}_3) + 1)}
{(N_{eq}^l({\bf p}) + 1)}
\label{eq:9i}
\end{equation}
\[
\times\Bigl(\vert{\cal T}_1\vert^2\Bigl\{
{\cal W}^l_{\bf p} -\,\frac{1}{2}\,{\cal W}^l_{{\bf p}_1} -
\frac{1}{4}\,({\cal W}^l_{{\bf p}_2} + {\cal W}^l_{{\bf p}_3})\Bigr\} +
\vert{\cal T}_2\vert^2\Bigl\{
{\cal W}^l_{\bf p} -\,\frac{1}{2}\,{\cal W}^l_{{\bf p}_3} -
\frac{1}{4}\,({\cal W}^l_{{\bf p}_1} + {\cal W}^l_{{\bf p}_2})\Bigr\}
\]
\[
+\,{\rm Re}\,({\cal T}_1{\cal T}_2^{\ast})\,\Bigl\{
{\cal W}^l_{\bf p} -\,\frac{1}{2}\,({\cal W}^l_{{\bf p}_1}
+ {\cal W}^l_{{\bf p}_3})\Bigr\} \Bigr).
\]
On the right-hand side of Eq.\,(\ref{eq:9i}) we introduce for brevity a notation
\begin{equation}
{\cal T}_1\equiv\frac{3}{2}\,g^2N_c
\left(\frac{{\rm Z}_l({\bf p})}{2\omega_{\bf p}^l\bar{u}^2(p)}\right)
\prod_{i=1}^{3}
\left(\frac{{\rm Z}_l({\bf p}_i)}{2\omega_{{\bf p}_i}^l\bar{u}^2(p_i)}\right)
\!\,^{\ast} \tilde{\Gamma} (p,p_1, -p_3, - p_2).
\label{eq:9o}
\end{equation}
Here, the function $\,^{\ast}\tilde{\Gamma}$ is defined by Eq.\,(\ref{eq:5a}),
and a function ${\cal T}_2$ is defined from ${\cal T}_1$ by a replacement
\[
\,^{\ast} \tilde{\Gamma} (p,p_1, -p_3, - p_2)\rightarrow
\,^{\ast} \tilde{\Gamma} (p,-p_2, p_1, - p_3).
\]
In deriving (\ref{eq:9i}) we use the following identities:
\[
{\rm tr}(T^aT^bT^c) = \frac{i}{2}\,N_cf^{abc},\quad
(T^aT^b)\,{\rm tr}(T^aT^cT^b) = \frac{1}{4}\,N_c^2T^c,\quad
(T^aT^bT^cT^aT^b) = 0,\;{\rm etc}.
\]
Since the equilibrium plasmon number density is proportional to the identity,
the commutator terms on the righ-hand side of Eq.\,(\ref{eq:9w}) vanish. Therefore
if we assume that the off-equilibrium function $\delta N^l$ is
perturbative small, then a linearized Boltzmann equation (\ref{eq:9u}),
(\ref{eq:9i}) in a certain sense is exact.

The linearized collision term (\ref{eq:9i}) for color plasmons
has noteworthy structure (cp. collision term for colorless case
(\ref{eq:5o}), (\ref{eq:5p})).
The expressions in curly brackets behind squared moduli
$\vert{\cal T}_1\vert^2$ and $\vert{\cal T}_2\vert^2$ coincide
in a structure with a similar expression entering into a linearized collision
term for color fluctuations of number density of hard gluons in form which was
proposed by Blaizot and Iancu \cite{bla3}. Besides, here additional
"interference"
term ${\rm Re}\,({\cal T}_1{\cal T}_2^{\ast})$ appears. Going immediately after
an expression in curly braces does not coincide with none of known expressions.
In spite of some complexity of obtained expression for
$\delta{\rm C}\,[{\cal W}_{\bf p}^l]$ it provides remark internal symmetry,
directly connected with existence of relation (\ref{eq:5s}).
Let us rewrite this relation in terms of ${\cal T}$ functions (\ref{eq:9o}).
For this purpose we introduce a third function ${\cal T}_3$, that is defined from
${\cal T}_1$, by replacement
\[
^{\ast} \tilde{\Gamma}(p, p_1, - p_3, - p_2) \rightarrow
\,^{\ast} \tilde{\Gamma}(p, p_1, - p_2, - p_3).
\]
Then from (\ref{eq:5s}) it follows
\[
{\cal T}_1 + {\cal T}_2 + {\cal T}_3 = 0.
\]
Now we set, for example, a function ${\cal T}_2 = - {\cal T}_1 - {\cal T}_3$
and substitute it into expression in round brackets in integrand (\ref{eq:9i}).
After simple algebraic transformations we have
\begin{equation}
\vert{\cal T}_1\vert^2\Bigl\{
{\cal W}^l_{\bf p} -\,\frac{1}{2}\,{\cal W}^l_{{\bf p}_2} -
\frac{1}{4}\,({\cal W}^l_{{\bf p}_1} + {\cal W}^l_{{\bf p}_3})\Bigr\} +
\vert{\cal T}_3\vert^2\Bigl\{
{\cal W}^l_{\bf p} -\,\frac{1}{2}\,{\cal W}^l_{{\bf p}_3} -
\frac{1}{4}\,({\cal W}^l_{{\bf p}_1} + {\cal W}^l_{{\bf p}_2})\Bigr\}
\label{eq:9p}
\end{equation}
\[
+\,{\rm Re}\,({\cal T}_1{\cal T}_3^{\ast})\,\Bigl\{
{\cal W}^l_{\bf p} -\,\frac{1}{2}\,({\cal W}^l_{{\bf p}_2}
+ {\cal W}^l_{{\bf p}_3})\Bigr\}.
\]
The structure of the last expression with an accuracy of replacement of momenta
fully coincides with a structure of Eq.\,(\ref{eq:9i}). The
linearized collision integral (\ref{eq:9i})
may be rewritten in terms of functions ${\cal T}_2$ and ${\cal T}_3$ by setting
${\cal T}_1 = - {\cal T}_2 - {\cal T}_3$. We obtain similar expression
with another permutation of momenta. Adding such obtained three expressions
and dividing by 3, we result in an expression that is more symmetric relative to
the external soft momenta and will be used somewhat below
\[
{\cal Q}({\bf p},{\bf p}_1;{\bf p}_2,{\bf p}_3) \equiv \frac{1}{3}\Bigr(
\vert{\cal T}_1\vert^2\Bigl\{
2{\cal W}^l_{\bf p} -\,\frac{1}{2}\,{\cal W}^l_{{\bf p}_3} -
\frac{3}{4}\,({\cal W}^l_{{\bf p}_1} + {\cal W}^l_{{\bf p}_2})\Bigr\}
\]
\[
+\,\vert{\cal T}_2\vert^2\Bigl\{
2{\cal W}^l_{\bf p} -\,\frac{1}{2}\,{\cal W}^l_{{\bf p}_1} -
\frac{3}{4}\,({\cal W}^l_{{\bf p}_2} + {\cal W}^l_{{\bf p}_3})\Bigr\} +
\vert{\cal T}_3\vert^2\Bigl\{
2{\cal W}^l_{\bf p} -\,\frac{1}{2}\,{\cal W}^l_{{\bf p}_2} -
\frac{3}{4}\,({\cal W}^l_{{\bf p}_1} + {\cal W}^l_{{\bf p}_3})\Bigr\}
\]
\begin{equation}
+\,{\rm Re}\,({\cal T}_1{\cal T}_2^{\ast})\,\Bigl\{
{\cal W}^l_{\bf p} -\,\frac{1}{2}\,({\cal W}^l_{{\bf p}_1}
+ {\cal W}^l_{{\bf p}_3})\Bigr\}
+\,{\rm Re}\,({\cal T}_1{\cal T}_3^{\ast})\,\Bigl\{
{\cal W}^l_{\bf p} -\,\frac{1}{2}\,({\cal W}^l_{{\bf p}_2}
+ {\cal W}^l_{{\bf p}_3})\Bigr\}
\label{eq:9a}
\end{equation}
\[
+\,{\rm Re}\,({\cal T}_2{\cal T}_3^{\ast})\,\Bigl\{
{\cal W}^l_{\bf p} -\,\frac{1}{2}\,({\cal W}^l_{{\bf p}_1}
+ {\cal W}^l_{{\bf p}_2})\Bigr\}\Bigr).
\]

By direct calculations we show that the color
current induced by off-equilibrium fluctuations of the color plasmons
number density $\delta N_{\bf p}^{l\,ab}$, covariantly conserves. In terms of
functions ${\cal W}_{\bf p}^{l\,a}$, the expressions for components of color
current (\ref{eq:9y}) is rewritten in the form
\begin{equation}
j_{0}^{\rm plasm}(x)=-\,gT^a\!\!\int\!\frac{d{\bf p}}{(2\pi)^3}\biggr(
\frac{\partial N_{\rm eq}({\bf p})}{\partial\omega_{\bf p}^l}\biggr)
{\cal W}_{\bf p}^{l\,a},\quad
{\bf j}^{\rm plasm}(x)=-\,gT^a\!\!\int\!\frac{d{\bf p}}{(2\pi)^3}\biggr(
\frac{\partial N_{\rm eq}({\bf p})}{\partial{\bf p}}\biggr)
{\cal W}_{\bf p}^{l\,a}.
\label{eq:9s}
\end{equation}
By linearized Boltzmann equation (\ref{eq:9u}) we have
\begin{equation}
{\cal D}^{\mu}(x)j_{\mu}^{\rm plasm}(x)=
-\,\frac{1}{3}\beta gN_c^2\!\!
\int\!\!\frac{d{\bf p}}{(2\pi)^3}\!
\int\!\prod_{i=1}^{3}\frac{d{\bf p}_i}{(2\pi)^3}\,
(2\pi)^4\delta({\bf p}+{\bf p}_1-{\bf p}_2-{\bf p}_3)
\delta(\omega^l_{\bf p}+\omega^l_{{\bf p}_1}-\omega^l_{{\bf p}_2}-
\omega^l_{{\bf p}_3})
\label{eq:9d}
\end{equation}
\[
\times N_{eq}^l({\bf p})N_{eq}^l({\bf p}_1)
(1+N_{eq}^l({\bf p}_2))(1+N_{eq}^l({\bf p}_3))
{\cal Q}({\bf p},{\bf p}_1;{\bf p}_2,{\bf p}_3),
\]
where the function ${\cal Q}$ on the righ-hand side is defined by Eq.\,(\ref{eq:9a}).
Now we show that integral on the righ-hand side of the last expression
equals to zero. As a first step we symmetrize integrand in Eq.\,(\ref{eq:9d})
with respect to permutation ${\bf p}\leftrightarrow{\bf p}_1$. Using
definitions of functions  ${\cal T}_1,\,{\cal T}_2$ and ${\cal T}_3$ and property of
invariance of the function $^{\ast}\tilde{\Gamma}(p,-p_1,-p_2,-p_3)$
(Eq.\,(\ref{eq:5d})), when the momenta order is reversed, we obtain laws of
transformation of these functions with replacement
${\bf p}\leftrightarrow{\bf p}_1$
\begin{equation}
{\cal T}_1 \rightarrow {\cal T}_3 ,\quad
{\cal T}_2 \rightarrow {\cal T}_2 ,\quad
{\cal T}_3 \rightarrow {\cal T}_1.
\label{eq:9f}
\end{equation}
Furthermore, we replace ${\bf p}\leftrightarrow{\bf p}_1$ in expression
(\ref{eq:9a}) with regard to (\ref{eq:9f}). Adding (\ref{eq:9a}) with such
an obtained expression and dividing by 2, we lead to the expression symmetric over
replacement ${\bf p}\leftrightarrow{\bf p}_1$:
\[
\frac{1}{2}\,\Bigl(\,\frac{5}{4}\,\vert{\cal T}_1\vert^2+
\frac{3}{2}\,\vert{\cal T}_2\vert^2+\frac{5}{4}\,\vert{\cal T}_3\vert^2
+\,\frac{1}{2}\,{\rm Re}\,({\cal T}_1{\cal T}_2^{\ast})
+\,{\rm Re}\,({\cal T}_1{\cal T}_3^{\ast})
+\,\frac{1}{2}\,{\rm Re}\,({\cal T}_2{\cal T}_3^{\ast})\Bigr)
\]
\begin{equation}
\times\Big\{{\cal W}_{\bf p}^l+{\cal W}_{{\bf p}_1}^l
-{\cal W}_{{\bf p}_2}^l-{\cal W}_{{\bf p}_3}^l\Bigr\}.
\label{eq:9g}
\end{equation}
As we see, this expression is automatically symmetric and with respect to
permutation ${\bf p}_2 \leftrightarrow{\bf p}_3$ (here, transformation laws
of functions ${\cal T}$ coincide with ({\ref{eq:9f})).

Let us consider now a crossed symmetry ${\bf p}\leftrightarrow{\bf p}_2,
{\bf p}_1 \leftrightarrow{\bf p}_3$. The statistical factor in integrand
(\ref{eq:9d}) is symmetric by virtue of identity
\[
N_{eq}^l({\bf p})N_{eq}^l({\bf p}_1)
(1+N_{eq}^l({\bf p}_2))(1+N_{eq}^l({\bf p}_3))=
(1+N_{eq}^l({\bf p}))(1+N_{eq}^l({\bf p}_1))N_{eq}^l({\bf p}_2)
N_{eq}^l({\bf p}_3).
\]
The transformation laws of functions ${\cal T}_i$ are trivial in this case
\[
{\cal T}_1 \rightarrow {\cal T}_1,\quad
{\cal T}_2 \rightarrow {\cal T}_2,\quad
{\cal T}_3 \rightarrow {\cal T}_3.
\]
By using these transformation laws and expression (\ref{eq:9g}) we see
that integrand on the right-hand side of Eq.\,(\ref{eq:9d}) is odd for
replacements ${\bf p}\leftrightarrow{\bf p}_2,\,
{\bf p}_1\leftrightarrow{\bf p}_3$. By virtue of this fact the integral
equals zero and a color current (\ref{eq:9s}) is covariantly conserves.

Closing this section note that without rigorous transport theory for
color plasmons in hand here, we given no any estimations for the amplitude
of mean gauge field ${\cal A}_{\mu}$. We can only made proposal that for
sufficiently strong background field, a structure of the kinetic equation
(\ref{eq:9q}) can be more involved. The interaction matrix elements
${\rm M}^{a a_1 \ldots a_{2n+1}}$ (or effective amplitudes), for example,
can depend by itself in a highly nontrivial way on mean field. In other
words, they can contain insertions of external mean field, in general,
of arbitrary order\footnote{The usual HTL-amplitudes in the presence of
a background gauge field, proposed by Blaizot and Iancu in Ref.\,\cite{bla2}
can be an example of such a dependence.}. This leads to the fact that mean
gluon field will be ``involved" in the dynamics of color plasmons not only
through the left-hand side of the kinetic equation, but through a collision
term.\\

\section{\bf Conclusion}
\setcounter{equation}{0}

On the basis of pure gauge sector of the Blaizot-Iancu equations we have
obtained the transport equation for the colorless plasmons, taking into
account $(2n+2)$-plasmon decays. The algorithm of successive
calculation of the probabilities of the decay processes is proposed.
The limiting value of the plasmon occupation number
$N_{\bf p}^l\,(\sim 1/g^2)$ is defined, whereby the decay processes with
an arbitrary number of soft external legs give contributions of the same
order in the coupling constant to the change of $N_{\bf p}^l$, and transport
theory becomes essentially nonlinear. Here, it should be pointed
to certain difficulty of a principal character, which arises when we
going over from low excited state ($A_{\mu}(X)\sim\sqrt{g}T$) to highly
excited one ($A_{\mu}(X)\sim T$) of system.

The kinetic equation (\ref{eq:2w}) describes the processes of plasmon
scattering. However moreover, there is a collective plasmon interaction
resulting in frequency shift of plasmons
\[
\omega_{\bf p}^l \rightarrow \tilde{\omega}_{\bf p}^l = \omega_{\bf p}^l
+\triangle\omega_{\bf p}^l.
\]
Here, $\tilde{\omega}_{\bf p}^l$ is the frequency renormalized due to the
interaction. In the linear plasma theory, i.e. in the theory with
infinitesimal amplitudes of plasma oscillations, the frequency spectrum
represents the function of values $(T,\,g, \ldots)$, not dependent
on waves energy (or their occupation number). In the framework of nonlinear
theory, the nonlinear dispersion correction represents in general case a
functional of plasmon number density $N_{\bf p}^l$. To lowest nontrivial
order in $N_{\bf p}^l$ the nonlinear frequency shift of longitudinal
oscillations is defined by the formula
\[
\triangle\omega_{\bf p}^l = g^2N_c\!
\int\!\frac{d{\bf p}_1}{(2\pi)^3}N_{{\bf p}_1}^l
\biggl(\frac{{\rm Z}_l({\bf p})}{2\omega_{\bf p}^l}\biggr)
\biggl(\frac{{\rm Z}_l({\bf p}_1)}{2\omega_{{\bf p}_1}^l}\biggr)
\biggl[\,\frac{1}{\bar{u}^2(p)\bar{u}^2(p_1)}
{\rm Re}\,^{\ast}\tilde{\Gamma}(p,p_1,-p,-p_1)\biggl]_{\rm on-shell}
\]
\begin{equation}
+\, O((N_{\bf p}^l)^2)+\ldots,
\label{eq:10q}
\end{equation}
where the function $\,^{\ast}\tilde{\Gamma}(p,p_1,-p,-p_1)$ is defined by
Eq.\,(\ref{eq:5a}). At the soft momentum scale, for the thermal
fluctuations of a gauge field we have an estimation for the
correction (\ref{eq:10q})
\[
\triangle\omega_{\bf p}^l\sim g^2N_cT\ll\omega_{\bf p}^l.
\]
Such for low excited state the frequency shift is perturbative small. In a
limiting case of a highly excited state (6.4) from Eq.\,(\ref{eq:10q})
we have the following estimation
\[
\triangle\omega_{\bf p}^l\sim gN_cT\sim\omega_{\bf p}^l,
\]
i.e. the nonlinear frequency shift becomes of the same
order\footnote{In this case to be sure in the right-hand side of
Eq.\,(\ref{eq:10q}) all terms of higher power in $N_{\bf p}^l$ are of the
same order in $g$, as the first term (similar to expansion
of generalized rates (\ref{eq:2e})).}
as a linear part of a spectrum $\omega_{\bf p}^l$. The spectrum
broadening over frequency takes place, and for strong field $A_{\mu}\leq T$ a dependence
$I^l(p,x)$ on a frequency (which in this work is defined in the form of the
quasiparticle approximation (\ref{eq:4s})) has no connection with
$\delta$-function. Therefore, the kinetic equation in the form (\ref{eq:2w})
in this limiting case becomes in general inappropriate for description of
nonlinear dynamics of soft excitations, and here, it should be derived more
general equation (or system of equations) for initial Wigner function
$I^l(p,x)$, taking into account the spectrum broadening effects. The examples
of conctraction of such an equations for usual plasma can be found in
Ref.\,\cite{kad}. Nevertheless, we suppose that somewhat simplified
approach to kinetic description of highly excited state of hot gluon
plasma considered in this paper will be useful in at least qualitative
analysis of complicated dynamics of interaction of soft boson
excitations taking place in the extremal medium state.

At the end of this section we note that at present we study a posibility of
an alternative way of description of nonlinear plasmons dynamics based on
a Hamiltonian formalism. The cornerstone of this approach is a fundamental
fact that the equations describing a collisionless hot gluon plasma in
HTL-approximation, possess a Hamiltonian structure, which was proposed by
Nair, Blaizot and Iancu \cite{efr, bla2, ian1}.
The last circumstance enables us to developed (at least for low excited
states) independent method of derivation of kinetic equation for soft
gluon plasma excitations.  Within the framework of Hamiltonian approach,
matrix elements of $(2n+2)$-plasmon decays are obtained as result of a special
canonical transformation, simplifying the interaction part of the plasmon
Hamiltonian. The exact consideration of this approach will be subject
of a separate publication.

\section*{\bf Acknowledgments}
This work was supported by Grant INTAS (No.\,2000-15) and
Grant for Young Scientist of Russian Academy of Sciences (No.\,1999-80).
The authors are grateful Alexander N. Vall for valuable discussions.\\

\newpage
\[
{\rm FIGURES}
\]
\begin{itemize}
\item[FIG.\,1.] The matrix element for four-plasmon decay. The wave
lines denote soft quasiparticles (plasmons) and the blob stands for HTL
resummation.
\item[FIG.\,2.] The typical tree-level Feynman diagram for (2n+2)-plasmon
decay with amputate external legs.
\end{itemize}
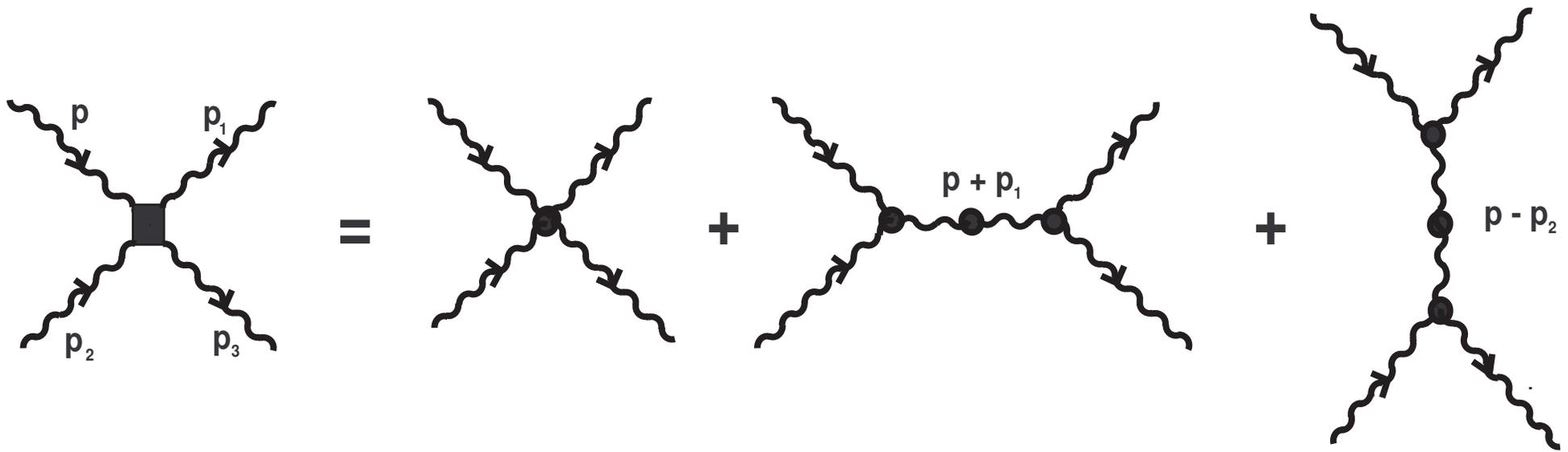
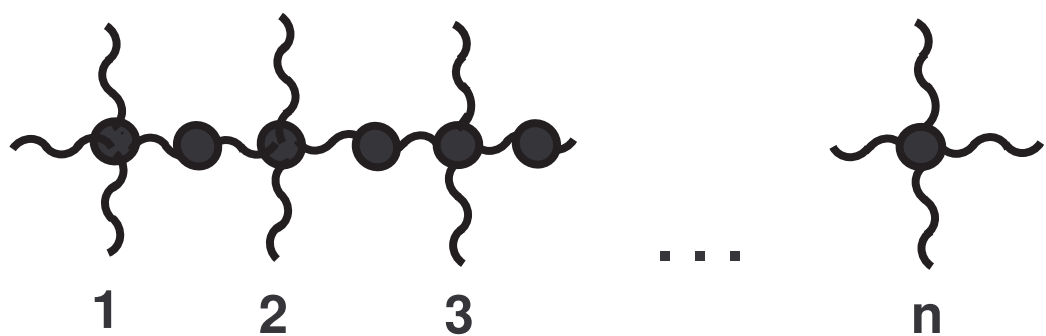
\end{document}